%% file: root.tex
\documentclass[letterpaper, 10 pt, conference]{ieeeconf}

\IEEEoverridecommandlockouts                              

\usepackage{graphics} 
\usepackage{epsfig} 
\usepackage{amsmath} 
\usepackage{amssymb}  
\usepackage[dvipsnames]{xcolor}
\usepackage{graphicx}  
\usepackage{subcaption}  
\usepackage{hyperref}

\usepackage{lipsum}
\usepackage{ulem}
\usepackage{siunitx}
\usepackage{graphics} 
\usepackage{epsfig} 
\usepackage{amsmath} 
\usepackage{amssymb}  
\DeclareMathAlphabet{\mathpzc}{OT1}{pzc}{m}{it}
\usepackage{graphicx}  
\usepackage{subcaption}  
\usepackage{booktabs}  
\usepackage{multirow}
\usepackage{tikz}
\usepackage{tikz-3dplot}
\usepackage[dvipsnames]{xcolor}
\newcommand{\mbf}[1]{\mathbf{#1}}
\newcommand{\sbf}[1]{\boldsymbol{#1}}
\newcommand{\mr}[1]{\mathring{#1}}
\newcommand{\T}[0]{\mathsf{T}}

\usepackage{soul}
\newif\ifhighlight
\highlightfalse

\ifhighlight
  \newcommand{\revised}[1]{\hl{#1}}  
  \newcommand{\revmargin}[1]{\marginnote{\tiny #1}} 
\else
  \newcommand{\revised}[1]{#1}       
  \newcommand{\revmargin}[1]{}       
\fi

\title{\LARGE \bf Equivariant Filter for Relative Attitude and Target's Angular Velocity Estimation}

\author{Gil Serrano, Bruno J. Guerreiro, Pedro Lourenço, and Rita Cunha 
\thanks{The work of Gil Serrano was supported by the PhD Grant PRT/BD/154275/2022 from MIT Portugal and Funda\c{c}{\~a}o para a Ci{\^e}ncia e a Tecnologia (FCT), Portugal [DOI: 10.54499/PRT/BD/154275/2022]. This work was also supported by FCT, Portugal through LARSyS [DOI: 10.54499/LA/P/0083/2020, 10.54499/UIDP/50009/2020, and 10.54499/UIDB/50009/2020] and project CAPTURE [DOI: 10.54499/PTDC/EEI-AUT/1732/2020]. 
}
\thanks{Gil Serrano, Bruno J. Guerreiro and Rita Cunha are with the Institute for Systems and Robotics, Instituto Superior Técnico, Universidade de Lisboa, Lisboa, Portugal.}
\thanks{Bruno J. Guerreiro is also with CTS/Uninova and LASI, School of Science and Technology at NOVA University Lisbon, Caparica, Portugal.}
\thanks{Pedro Lourenço is with the GNC Division, Flight Segment and Robotics, GMV, Lisboa, Portugal.}
\thanks{Corresponding author: Gil Serrano ({gil.serrano@tecnico.ulisboa.pt}).}
}

\begin{document}

\maketitle
\thispagestyle{empty}
\pagestyle{empty}

\input{sections/00_abstract}

\input{sections/01_introduction}

\input{sections/02_problem_statement}

\input{sections/03_equivariant_formulation}

\input{sections/04_filter_design}

\input{sections/05_simulation_results}

\input{sections/06_experimental_results}

\input{sections/07_conclusion}



\input{sections/99_appendix}

\section*{Ackowledgment}
G. Serrano thanks M. Jacinto for the insightful discussions and valuable perspectives.



\addtolength{\textheight}{-1.5cm}   

\bibliographystyle{IEEEtran}
\bibliography{IEEEabrv,mybibfile}

\end{document}

%% file: sections/00_abstract.tex
\begin{abstract}
Accurate estimation of the relative attitude and angular velocity between two rigid bodies is fundamental in aerospace applications such as spacecraft rendezvous and docking. In these scenarios, a chaser vehicle must determine the orientation and angular velocity of a target object using onboard sensors. 
This work addresses the challenge of designing an Equivariant Filter (EqF) that can reliably estimate both the relative attitude and the target's angular velocity using noisy observations of two known, non-collinear vectors fixed in the target frame. To derive the EqF, a symmetry for the system is proposed and an equivariant lift onto the symmetry group is calculated.
Observability and convergence properties are analyzed.
Simulations demonstrate the filter's performance, with Monte Carlo runs yielding statistically significant results. The impact of low-rate measurements is also examined and a strategy to mitigate this effect is proposed. Experimental results, using fiducial markers and both conventional and event cameras for measurement acquisition, further validate the approach, confirming its effectiveness in a realistic setting.
\end{abstract}

%% file: sections/01_introduction.tex
\section{INTRODUCTION}

In the past decade, there has been a growing interest in the development and validation of On-Orbit Servicing (OOS) and Active Debris Removal (ADR) technologies \cite{flores-abad_review_2014, biesbroek_edeorbit_2017, hatty_viability_2022}, driven by the ever-increasing number of satellites deployed each year. This rapid rise in satellite deployments multiplies the amount of space debris in strategic orbits and negatively impacts the operational lifetime of satellites \cite{ailor_effect_2017}.

In OOS and ADR missions, the chaser or servicer spacecraft will need to approach the target spacecraft and synchronize its motion before performing the planned operations. As such, the chaser must estimate the relative attitude and angular velocity of the target, as well as the relative position and linear velocity. To do so, the chaser can rely on data acquired via electro-optical sensors, such as conventional or event cameras and light detection and range (LiDAR), by tracking features \cite{pesce_stereovision-based_2017,capuano_monocular-based_2020} or fiducial markers \cite{sansone_relative_2018, vela_pose_2022} on the target. 

Recently, learning-based approaches to relative pose estimation of an uncooperative target spacecraft have become increasingly popular \cite{kisantal_satellite_2020,sharma_neural_2020,kaidanovic_deeplearning_2022,bechini_robust_cnn_2024}. 
Despite the current interest, these methods come with a variety of drawbacks, especially when considering end-to-end relative attitude or pose estimation with neural networks. Besides being computationally demanding, these networks currently rely heavily on visual data obtained from simulations, meaning that the models suffer from the sim-to-real gap in the transfer to real data. Sensor noise, lighting shifts, or partial occlusions not considered in simulation might hinder the performance of these networks in real scenarios. Furthermore, these models provide no stability proofs or convergence guarantees and could have problems estimating states on non-Euclidean manifolds. Nevertheless, learning-based approaches could be useful in perception, an area where these methods have had a significant impact and outperform classical computer vision algorithms, or as sources of pose measurements that are fed to a filter, as is done in {\cite{park_adaptive_ukf_2023}}.

Filtering techniques, especially those based on Kalman filtering, are ubiquitous in relative motion estimation and the Multiplicative Extended Kalman Filter (MEKF) has been the the industry standard for attitude estimation in space applications {\cite{crassidis_survey_2007}}. 
The MEKF state consists of a nominal state (unit quaternion) for attitude and a separate error state. At each filter update, this error is turned into a quaternion and multiplied by the nominal state. This operation maintains the attitude quaternion estimate with unit norm. In contrast, the Extended Kalman Filter (EKF) updates the quaternion additively, which breaks the unit norm constraint and requires correction, making the MEKF more geometrically consistent.
%
In {\cite{pesce_comparison_2019}}, Pesce et al. compare the performance of a MEKF, a minimum-energy filter on SO(3) {\cite{zamani_minimum_energy_2013}}, a constant-gain attitude observer, and a second-order minimum-energy filter {\cite{saccon_second_order_optimal_2016}} in relative attitude estimation of uncooperative space objects. The study assumes that pose measurements are attained through a monocular camera system as output of a model-matching algorithm, similar to {\cite{park_adaptive_ukf_2023}}, and runs Monte Carlo simulations with varying degrees of uncertainty in the measurements and parameters. Based on the outcome of all tested conditions, the authors conclude that, in general, the MEKF presents the fastest convergence, but shows inferior performance when compared to the minimum-energy filters and the constant-gain observer in the steady-state. Despite the thorough analysis, this work only analyzed relative attitude estimation and, in the scenario considered in the present work, we are interested in estimating the target's angular velocity as well. 

Several Kalman filter–based methods have been proposed to estimate the relative state of an uncooperative target, including its relative attitude and angular velocity.
In {\cite{pesce_stereovision-based_2017}}, an Iterative Extended Kalman Filter (IEKF) is proposed to estimate the relative pose, linear and angular velocities, and the moment of inertia ratios of an uncooperative target, using a stereo-vision system to detect and track features on the target. The algorithm is validated by numerical simulations that verify accuracy and robustness in different settings and initial conditions.
An approach based on the Unscented Kalman Filter (UKF) is presented in {\cite{wang_stereovision_ukf_2019}} to estimate the relative states and moment of inertia ratios of an uncooperative spacecraft, also using stereo-vision measurements. The method uses kinematic constraints and optimization strategies to have better initial conditions for the filter and is shown to have improved robustness to large measurement noise.
In {\cite{pesce_autonomous_2019}}, a complete relative navigation strategy using monocular measurements is proposed. The rotational motion estimation is handled by an adapted version of the filter proposed in {\cite{saccon_second_order_optimal_2016}}. In their model, the authors assume that the derivative of the relative angular velocity is zero, to increase the capability of handling unknown or partially unknown objects, by avoiding the dependence on the inertia tensor of the target spacecraft. Several scenarios are analyzed and, though satisfactory results for position and attitude estimation are obtained, the authors conclude that a higher relative angular velocity implies a higher error and a slower convergence.
Capuano et al. present in {\cite{capuano_monocular-based_2020}} a multi-stage approach, using both monocular vision and LiDAR. The proposed architecture first estimates the relative angular velocity by solving an algebraic optimization problem based on the optical flow. Then, an EKF is implemented to estimate the relative position, linear velocity, and attitude. The rationale for this division is that removing the relative angular velocity and inertia parameters from the EKF reduces the filter state dimension, thereby decreasing the computational burden.
Another multi-stage solution is introduced by Nocerino et al. {\cite{nocerino_lidar_ukf_2021}} to estimate the inertia parameters and the relative state of an unknown target, using LiDAR measurements. The moments of inertia ratios are first computed via an iterative procedure and then a UKF is used to determine the entire relative state.
In \cite{zivan_dual_2022}, an EKF is designed for spacecraft relative pose estimation. The relative pose is represented by a dual quaternion, although the geometric constraints imposed on the state by the dual quaternion are not handled directly, nor exploited, but rather enforced in an \textit{ad hoc} manner. 
Barbier and Gao \cite{barbier_relative_2023} present an EKF-based method for relative pose estimation in a scenario with an unknown uncooperative target. Assuming feature detection and tracking, a cascade of Error-State Kalman Filters (ESKF) first estimates the chaser's trajectory and attitude, along with the target's shape, and then tracks the position and attitude of the target. The filter is validated in simulation, though no formal analysis is offered.
%
In \cite{parreira_pose_2024}, Parreira et al. propose an eXogenous Kalman Filter (XKF) to estimate the relative pose, linear and angular velocities, and moment-of-inertia ratios of uncooperative targets. An XKF is a two-stage nonlinear estimator, composed of a globally convergent observer cascaded with an EKF-like filter \cite{johansen_fossen_xkf_2017}. The authors conclude that, given sufficiently exciting target rotational motion, this filter offers convergence guarantees and near-optimal performance. Monte Carlo simulations are used to evaluate the filter's performance, which is observed to be similar to that of an EKF. 

In this paper, we are interested in the estimation of the relative attitude between the chaser and the target spacecraft and the target's angular velocity, expressed in the chaser frame, and propose an Equivariant Filter (EqF) to this effect.
This type of filter is based on the equivariant systems theory developed by Mahony et al. \cite{mahony_observers_2013, mahony_equivariant_2020, mahony_observer_2022}. Building on the idea that some systems' states lie on a smooth manifold with a transitive Lie group symmetry, the EqF works by embedding the observer state within the symmetry group, linearizing the global error dynamics derived from the system's equivariance, and applying principles from EKF design \cite{mahony_equivariant_2021, van_goor_equivariant_2023}. 
Unlike more traditional filters that might overlook the underlying geometry of the state space, the EqF leverages the inherent symmetry of the system and takes advantage of equivariance to obtain more consistent and robust estimates.
The EqF can be applied to any system that exhibits the equivariance property, without requiring the system model to be described specifically on the Lie group.

Equivariant observers have been applied to a number of estimation problems, including attitude estimation and pose estimation. In \cite{ng_attitude_2019}, Ng et al. design an equivariant observer for second-order attitude kinematics. Using measurements from accelerometers and magnetometers, the observer is able to estimate the attitude and angular velocity of a rigid body, proving almost global asymptotic stability and local uniform exponential stability of the estimation error. 
The authors extend the work in \cite{ng_equivariant_2020}, proposing an equivariant observer for second order pose estimation of a rigid body. The observer uses accelerometer and gyroscope measurements, as well as partial pose measurements, and exploits the second-order kinematic model and its symmetry group, yielding asymptotic convergence.
In \cite{fornasier_overcoming_2022}, Fornasier et al. propose a symmetry and derive an EqF for gyroscope-aided attitude estimation, using direction measurements, which also estimated the bias of the gyroscope. The estimator improved the transient response and the asymptotic bias estimation compared to other state-of-the-art approaches.
To derive an EqF for biased inertial-based navigation, the authors present in \cite{fornasier_equivariant_2022} a novel equivariant lift for the system onto the symmetry group and derive a filter that simultaneously estimates the navigation state and the input measurement biases. Simulated and real-world data demonstrate the increased robustness of the filter in a broad set of initial conditions and its improved accuracy when compared to a MEKF approach.

The main contributions of this work with respect to the related literature are: 
i) the formulation and exploitation of an underlying Lie group symmetry and equivariant lift for the combined estimation of relative attitude and target's angular velocity by a geometrically consistent, symmetry-preserving filter, using vector measurements that correspond to physical features of the target;
ii) the analysis of the filter’s behavior with low measurement rates and the proposal of a strategy to mitigate performance degradation;
iii) the validation of the filter on physical hardware, using both conventional and event cameras for data acquisition, thus demonstrating practical robustness with modern sensors.

The remainder of this paper is organized as follows: Section\;\ref{sec:problem_statement} states the problem of estimating the relative attitude between two frames and the angular velocity of a target frame; Section\;\ref{sec:equivariant_formulation} explains the the symmetry of the system and the equivariant lift onto the Lie group; Section\;\ref{sec:filter_design} describes the Equivariant Filter's derivation and convergence analysis; Section\;\ref{sec:simulation_results} presents simulation results, compares the approach with an EKF and studies the effect of low-rate measurements; Section\;\ref{sec:experimental_results} shows experimental results with data acquired from a conventional and an event camera; and finally, Section\;\ref{sec:conclusion} offers concluding remarks.

%% file: sections/02_problem_statement.tex
\section{PROBLEM STATEMENT} 
\label{sec:problem_statement}

We define the relevant reference frames and present the kinematic system that is going to be studied, as well as the assumed measurement model and perform the observability analysis. Additionally, we introduce the symmetry group that will be used to lift the system and the operations that will be necessary for the definition of the filter equations.

\subsection{Reference Frames} 
\label{subsec:reference_frames}

\begin{figure}
    \centering
    \includegraphics[width=\columnwidth]{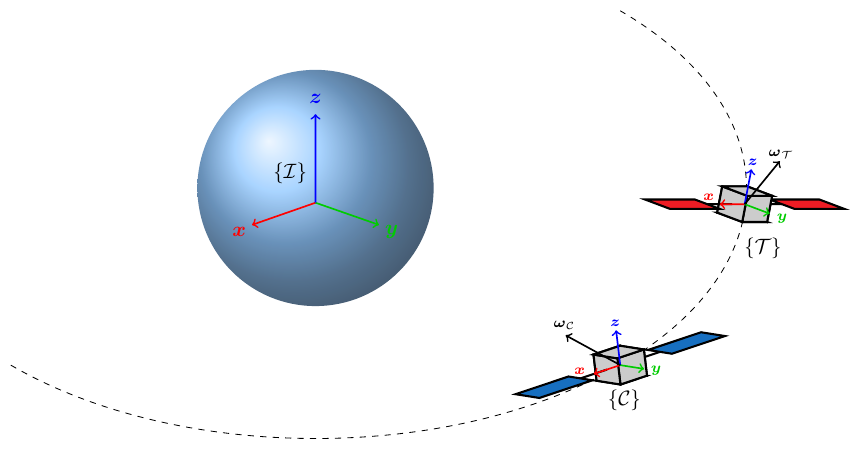}
    \caption{Chaser and target scenario with relevant reference frames.}
    \label{fig:reference_coordinate_frames}
\end{figure}

Consider a system with two rotating body-fixed frames, defined as a target frame $\{\mathcal{T}\}$ and a chaser frame $\{\mathcal{C}\}$, and an inertial frame $\{\mathcal{I}\}$. The rotation matrices ${}^\mathcal{I}_\mathcal{T}\mbf{R}$ and ${}^\mathcal{I}_\mathcal{C}\mbf{R}$ represent the attitude of the target and chaser frames with respect to the inertial frame, respectively.
We denote the angular velocities of the target and chaser, with respect to the inertial frame, expressed in their respective frames, by $\sbf{\omega}_\mathcal{T}$ and $\sbf{\omega}_\mathcal{C}$. In Fig.\,{\ref{fig:reference_coordinate_frames}}, we provide a representation of the considered scenario with the reference frames defined above.


\subsection{Kinematic System} 
\label{subsec:kinematic_system}

The attitude kinematics of the target and chaser, relative to the inertial frame, are expressed by
\begin{align}
{}^\mathcal{I}_\mathcal{T}\dot{\mbf{R}} \,&=\, {}^\mathcal{I}_\mathcal{T}\mbf{R}\,{\bigl(\sbf{\omega}_\mathcal{T}\bigr)}^{\wedge}\,,\\
{}^\mathcal{I}_\mathcal{C}\dot{\mbf{R}} \,&=\, {}^\mathcal{I}_\mathcal{C}\mbf{R}\,{(\sbf{\omega}_\mathcal{C}\bigr)}^{\wedge}\,,
\end{align}
where the notation $(\mbf{x})^{\wedge}$, for $\mbf{x}\in\mathbb{R}^3$, represents the skew-symmetric matrix constructed from the vector $\mbf{x}$, such that $(\mbf{x})^{\wedge}\mbf{y} = \mbf{x}\times\mbf{y},\,\forall\mbf{y}\in\mathbb{R}^3$. The inverse operation to ${(\cdot)^{\wedge}}$ is denoted as ${(\cdot)^{\vee}}$, such that ${{((\mbf{x})^{\wedge})}^{\vee} = \mbf{x},\,\forall\mbf{x}\in\mathbb{R}^3}$.

The relative attitude between the chaser frame $\{\mathcal{C}\}$ and the target frame $\{\mathcal{T}\}$ is represented by ${{}^\mathcal{T}_\mathcal{C}\mbf{R} = {{}^\mathcal{I}_\mathcal{T}\mbf{R}^\T} \,{}^\mathcal{I}_\mathcal{C}\mbf{R}}$. Thus, the relative attitude kinematics are given by
\begin{equation}
        {}^\mathcal{T}_\mathcal{C}\dot{\mbf{R}} \,=\, {}^\mathcal{T}_\mathcal{C}\mbf{R}\, {\bigl(\sbf{\omega}_\mathcal{C}\bigr)}^{\wedge} - {\bigl(\sbf{\omega}_\mathcal{T}\bigr)}^{\wedge}\, {}^\mathcal{T}_\mathcal{C}\mbf{R}\,.
        \label{eq:R_dot_notation_1}
\end{equation}
The above system can be rewritten as
\begin{equation}
\begin{split}
    {}^\mathcal{T}_\mathcal{C}\dot{\mbf{R}} \,&=\, {}^\mathcal{T}_\mathcal{C}\mbf{R}\, \Bigl({\bigl(\sbf{\omega}_\mathcal{C}\bigr)}^{\wedge} - {{}^\mathcal{T}_\mathcal{C}\mbf{R}^\T}{\bigl(\sbf{\omega}_\mathcal{T}\bigr)}^{\wedge}\, {}^\mathcal{T}_\mathcal{C}\mbf{R}\Bigr)\\
    &=\, {}^\mathcal{T}_\mathcal{C}\mbf{R}\, \Bigl({\bigl(\sbf{\omega}_\mathcal{C}\bigr)}^{\wedge} - {({{}^\mathcal{T}_\mathcal{C}\mbf{R}^\T}\,\sbf{\omega}_\mathcal{T}})^{\wedge}\Bigr)\,.
\end{split}
\label{eq:R_dot_notation_2}
\end{equation}
To simplify the notation, let us define ${\mbf{R}\,\equiv\,{}^\mathcal{T}_\mathcal{C}\mbf{R}}$, ${\mbf{u}\,\equiv\,\sbf{\omega}_\mathcal{C}}$, and ${\sbf{\omega}\,\equiv\,{{}^\mathcal{T}_\mathcal{C}\mbf{R}^\T}\,\sbf{\omega}_\mathcal{T}}$, the target's angular velocity expressed in the chaser frame, resulting in the system
\begin{equation}
    \dot{\mbf{R}} = \mbf{R}(\mbf{u} - \sbf{\omega})^{\wedge}\,.
    \label{eq:R_dot_simplified_notation}
\end{equation}

We are interested in studying the case where the chaser's angular velocity, $\mbf{u}$, is known and the target's angular velocity is unknown but constant, i.e., $\dot{\sbf{\omega}}_\mathcal{T} = \mbf{0}$, which is consistent with a body rotating about the principal axis with the largest moment of inertia. Nevertheless, we will consider for now a more general system, where ${\dot{\sbf{\omega}}_\mathcal{T} = \mbf{a}_\mathcal{T}}$, with $\mbf{a}_\mathcal{T}$ being the target's angular acceleration expressed in the target frame. The system of interest can be recovered by simply setting ${\mbf{a}_\mathcal{T} = \mbf{0}}$. The expression for ${\dot{\sbf{\omega}}}$ is
\begin{equation}
    \begin{split}
        \dot{\sbf{\omega}} &= \dot{\mbf{R}}^\T\sbf{\omega}_\mathcal{T} + \mbf{R}^\T\dot{\sbf{\omega}}_\mathcal{T}\\
        &= -(\mbf{u})^{\wedge}\mbf{R}^\T\sbf{\omega}_\mathcal{T} + \mbf{R}^\T(\sbf{\omega}_\mathcal{T})^{\wedge}\,\sbf{\omega}_\mathcal{T} + \mbf{R}^\T\mbf{a}_\mathcal{T}\\
        &= \sbf{\omega}\times\mbf{u} +\mbf{a}\,,\\
    \end{split}
    \label{eq:omega_dot_derivation}
\end{equation}
where ${\mbf{a} = \mbf{R}^\T\mbf{a}_\mathcal{T}}$ is the target's angular acceleration, expressed in the chaser frame.

We formulate a state ${(\mbf{R},\,\sbf{\omega}) \in \mathcal{M}  = \mathcal{SO}(3) \times \mathbb{R}^3}$, ${\mathcal{SO}(3)}$ the $SO(3)$-torsor \cite{mahony_observers_2013}, $SO(3)$ the special orthogonal group, and input ${(\mbf{u},\mbf{a}) \in \mathbb{L} = \mathbb{R}^3\times \mathbb{R}^3}$, where $\mathcal{M}$ and $\mathbb{L}$ are the state and input manifolds, respectively. Thus, the complete system under consideration is ${(\dot{\mbf{R}}, \dot{\sbf{\omega}}) = f((\mbf{R},\sbf{\omega}), (\mbf{u},\mbf{a}))}$, as described by
\begin{align}
    \dot{\mbf{R}} &= \mbf{R}(\mbf{u} - \sbf{\omega})^\wedge\,, \label{eq:kinematic_system_R} \\
    \dot{\sbf{\omega}} &= \sbf{\omega}\times\mbf{u} + \mbf{a}\,.\label{eq:kinematic_system_omega}
\end{align}

\subsection{Measurement Model} 
\label{subsec:measurement_model}

In this scenario, we assume that there are two non-collinear reference vectors, which are constant when expressed in the frame $\{\mathcal{T}\}$, denoted by $\mr{\mbf{d}}_1$, $\mr{\mbf{d}}_2\in\mathbb{R}^3$. We represent these vectors as $\mbf{d}_1$ and $\mbf{d}_2$, when expressed in the frame $\{\mathcal{C}\}$, with each being given by
\begin{equation}
    \mbf{d}_i = \mbf{R}^\T \mathring{\mbf{d}}_i,\,i = \{1,2\}\,.
    \label{eq:measurement_direction}
\end{equation}
These vectors could be inferred from known characteristics of the target satellite's surface or measured using fiducial markers. 
As such, the output of the system, represented by ${\mbf{y} = h(\mbf{R}, \sbf{\omega})}$, where $h:\mathcal{M}\to\mathcal{Y}$ and ${\mathcal{Y}=\mathbb{R}^3\times\mathbb{R}^3}$ denotes the output space, is given by
\begin{equation}
    \mbf{y} = h(\mbf{R}, \sbf{\omega}) = (\mbf{d}_1,\, \mbf{d}_2).
    \label{eq:system_output_y}
\end{equation}
Alternatively, the output can also be expressed in vector form as ${\mbf{y} = [\mbf{d}_1\,\, \mbf{d}_2]^\T\in\mathbb{R}^{6}}$.

In reality, the measurements will be noisy. We assume that the measured vectors are corrupted by noise as decribed by
\begin{equation}
    \mbf{d}_i^n = \sbf{R}(\theta, \mbf{n})\,\mbf{d}_i
\label{eq:measurement_direction_noisy}
\end{equation}
where ${\sbf{R}(\theta, \mbf{n}):\mathbb{R}\times\mathcal{S}^2\to \mathcal{SO}(3)}$ denotes a rotation matrix that encodes a rotation by angle $\theta$ about an axis $\mbf{n}$. The rotation angle is assumed to follow a zero-mean normal distribution, that is, ${\theta\sim\mathpzc{N}(0, {\sigma_{\theta}}^2)}$, where $\sigma_{\theta}$ is the standard deviation. The rotation axis follows a uniform distribution on $\mathcal{S}^2$, $\mbf{n}\sim\mathpzc{U}(\mathcal{S}^2)$. This is achieved by taking a vector ${\bar{\mbf{n}}\sim\mathpzc{N}(0, {\sigma_{\bar{\mbf{n}}}}^2\mbf{I})}$, where ${\sigma_{\bar{\mbf{n}}}}$ is the standard deviation in each dimension, and normalizing it such that $\mbf{n} = \bar{\mbf{n}}/\|\bar{\mbf{n}}\|$. As a result of this process, ${\sigma_{\bar{\mbf{n}}}}$ does not influence the distribution of the normalized vector. As such, the noisy measurements are small deviations from the true vectors.

\subsection{Observability Analysis}
\label{subsec:observability_analysis}

Let us confirm that the system is observable, i.e., that the relative attitude and the target's angular velocity can effectively be estimated, given the aforementioned measurements. We follow the same approach as in \cite{krener_nonlinear_1977,nijmeijer_nonlinear_1990}, where a nonlinear observability rank condition is used. 

To do so, we first vectorize the attitude rotation matrix ${\mbf{r} = \text{vect}(\mbf{R}) \in \mathbb{R}^9}$, and define a vector state ${\mbf{x} = [\mbf{r}^\T\;\sbf{\omega}^\T]^\T}\in\mathbb{R}^{12}$. Then, the observability matrix is constructed as
\begin{equation}
    \mathcal{O}(\mbf{x}) = \begin{bmatrix}
        L_f^0 h(\mbf{x})\\
        L_f^1 h(\mbf{x})\\
        L_f^2 h(\mbf{x})\\
        \vdots
    \end{bmatrix}\,,
    \label{eq:observability_matrix}
\end{equation}
where ${L_f^0 h = h(\mbf{x})}$ and ${L_f^n h = \frac{\partial L_f^{n-1} h}{\partial \mbf{x}}\cdot f(\mbf{x})}$ is the $n$\textsuperscript{th}-degree Lie derivative of the measurement function with respect to the system function $f$. If all states of the system are observable, then ${\text{rank}\left(\frac{d\mathcal{O}}{d\mbf{x}}\right)} = 12$. For this system and observability matrix, we have that ${\text{rank}\left(\frac{d\mathcal{O}}{d\mbf{x}}\right)} = 9$, with the unobserved states ${\mbf{x}_{\text{unobs}}}$ being
\begin{equation}
    \mbf{x}_{\text{unobs}} = \mbf{R}^\T \left( \mr{\mbf{d}}_1 \times \mr{\mbf{d}}_2 \right)\,.
\end{equation}

However, since we know that ${\mbf{R}\mbf{R}^\T = \mbf{I}}$, we can add some fictitious measurements to $h$ that reflect this constraint. We consider an expanded measurement model ${\bar{h}:\mathcal{M}\to\mathcal{\bar{Y}}}$, given by
\begin{equation}
    \bar{h} = \bigl( \mbf{R}^\T \mr{\mbf{d}}_1,\, \mbf{R}^\T \mr{\mbf{d}}_1,\,\mbf{R}\mbf{R}^\T \mbf{e}_1,\,\mbf{R}\mbf{R}^\T \mbf{e}_2,\,\mbf{R}\mbf{R}^\T \mbf{e}_3 \bigr)\,,
    \label{eq:expanded_measurement_model}
\end{equation}
where $\mbf{e}_1,\,\mbf{e}_2,\,\mbf{e}_3$ are the standard basis vectors of $\mathbb{R}^3$. These added fictitious measurements would give each of the columns of the identity matrix. Reconstructing the observability matrix \eqref{eq:observability_matrix} with the expanded measurement model, we get ${\text{rank}\left(\frac{d\mathcal{O}}{d\mbf{x}}\right)} = 12$. Thus, taking into account its topological properties, we conclude that the system is observable. 


\subsection{Symmetry Group} 
\label{subsec:symmetry_group}

The symmetry group that is considered for this problem is the Lie group $G = SE(3)$, the special Euclidean group, with Lie-algebra $\mathfrak{g} = \mathfrak{se}(3)$. 

An element of the group is represented as ${(\mbf{Q},\mbf{q})\in SE(3)}$, where ${\mbf{Q}\in SO(3)}$ is a rotation matrix and ${\mbf{q}\in\mathbb{R}^3}$ is a 3-dimensional vector. The identity is ${(\mbf{I}, \mbf{0})}$ and the inverse element is ${(\mbf{Q}, \mbf{q})^{-1} = (\mbf{Q}^\T, -\mbf{Q}^\T\mbf{q})}$. The group multiplication in ${SE(3)}$ is given by
\begin{equation}
    (\mbf{Q}_2,\mbf{q}_2)\cdot(\mbf{Q}_1,\mbf{q}_1) = (\mbf{Q}_2\mbf{Q}_1, \mbf{Q}_2\mbf{q}_1 + \mbf{q}_2)
\end{equation}
for any two elements $(\mbf{Q}_1,\mbf{q}_1)$, $(\mbf{Q}_2,\mbf{q}_2) \in SE(3)$.
Using the definition of group multiplication, the left and right multiplication maps are respectively described by 
\begin{align}
{L_{(\mbf{Q}_1,\mbf{q}_1)}(\mbf{Q}_2,\mbf{q}_2)} &= (\mbf{Q}_1,\mbf{q}_1)\cdot(\mbf{Q}_2,\mbf{q}_2)\,, \label{eq:left_group_multiplication}\\
{R_{(\mbf{Q}_1,\mbf{q}_1)}(\mbf{Q}_2,\mbf{q}_2)} &= (\mbf{Q}_2,\mbf{q}_2)\cdot (\mbf{Q}_1,\mbf{q}_1)\,.\label{eq:right_group_multiplication}
\end{align}

An element of the Lie-algebra is represented as $(\mbf{S}, \mbf{s})\in\mathfrak{se}(3)$, where $\mbf{S}\in \mathfrak{so}(3)$ is a skew-symmetric matrix and $\mbf{s}\in\mathbb{R}^3$ is a 3-dimensional vector.

The differentials of the left and right multiplication maps, $DL_{(\mbf{Q},\mbf{q})}(\mbf{S}, \mbf{s})$ and $DR_{(\mbf{Q},\mbf{q})}(\mbf{S}, \mbf{s})$, describe how an element of the Lie-algebra $(\mbf{S}, \mbf{s})$ is transformed under the action of the left and right multiplications by a group element $(\mbf{Q}, \mbf{q})$. These operations are given by
\begin{align}
    DL_{(\mbf{Q},\mbf{q})}(\mbf{S}, \mbf{s}) &= (\mbf{Q} \mbf{S},\, \mbf{Q} \mbf{s})\,, \label{eq:left_group_multiplication_differential}\\
    DR_{(\mbf{Q},\mbf{q})}(\mbf{S}, \mbf{s}) &= (\mbf{S}\mbf{Q} ,\, \mbf{S}\mbf{q} + \mbf{s})\,. \label{eq:right_group_multiplication_differential}
\end{align}
Having defined the differentials of the left and right group multiplications, we define the adjoint ${\text{Ad}_{(\mbf{Q},\mbf{q})}(\mbf{S},\mbf{s})}$ as
\begin{equation}
\begin{split}
    \text{Ad}_{(\mbf{Q},\mbf{q})}(\mbf{S},\mbf{s}) &= DL_{(\mbf{Q},\mbf{q})} \cdot DR_{(\mbf{Q},\mbf{q})^{-1}} (\mbf{S},\mbf{s})\\
    &= (\mbf{Q}\mbf{S}\mbf{Q}^\T, -\mbf{Q}\mbf{S}\mbf{Q}^\T\mbf{q} + \mbf{Q}\mbf{s})\,.
\end{split}
\label{eq:group_adjoint}
\end{equation}

%% file: sections/03_equivariant_formulation.tex
\section{EQUIVARIANT SYSTEM FORMULATION}
\label{sec:equivariant_formulation}

To design an Equivariant Filter, using the previously defined group symmetry, we must deduce the state and input group actions, set an origin for the coordinate system on the manifold, verify if the system's output is equivariant, and finally derive an equivariant lift of the system onto the group.
\subsection{System Extension}
\label{subsec:system_extension}

To prove equivariance, we will extend the system by adding virtual inputs, a strategy followed in \cite{ng_attitude_2019, ng_equivariant_2020} for other second-order systems. Due to the differential equation related to the angular velocity \eqref{eq:kinematic_system_omega}, the system requires the addition of two virtual inputs ${(\mbf{v}, \mbf{w})\in\mathbb{R}^3\times \mathbb{R}^3}$. As such, the extended system is 
\begin{align}
    \dot{\mbf{R}} &= \mbf{R}(\mbf{u} - \sbf{\omega} + \mbf{v})^{\wedge}\,,\\
    \dot{\sbf{\omega}} &= (\sbf{\omega} + \mbf{w}) \times \mbf{u} + \mbf{a}\,,
\end{align}
which we denote by $(\dot{\mbf{R}}, \dot{\sbf{\omega}}) = \Bar{f}((\mbf{R},\sbf{\omega}), (\mbf{u},\mbf{a},\mbf{v},\mbf{w}))$. The original system is recovered by setting $\mbf{v} = \mbf{w} = \mbf{0}$. The input manifold is now ${\mathbb{L} = \mathbb{R}^3 \times \mathbb{R}^3 \times \mathbb{R}^3\times \mathbb{R}^3}$.

\subsection{Right State Action}
\label{subsec:right_state_action}
As introduced in Section\;\ref{subsec:symmetry_group}, the symmetry group for the system is $G=SE(3)$ and we take the group variable $(\mbf{Q}, \mbf{q}) \in G$. We propose a map ${\phi: G \times \mathcal{M}\to \mathcal{M}}$, which is a right group action on the manifold. For the system at hand, this function is
\begin{equation}
    \phi((\mbf{Q},\mbf{q}), (\mbf{R},\sbf{\omega})) = (\mbf{R}\mbf{Q}, \mbf{Q}^\T(\sbf{\omega} - \mbf{q}))\,.
    \label{eq:state_right_group_action}
\end{equation}

\subsection{Coordinate System's Origin}
\label{subsec:coordinate_system_origin}
We need to define an origin $(\mathring{\mbf{R}},\mathring{\sbf{\omega}})$ for our coordinate system, which will be used to translate the filter's estimates from the group back to the state manifold. Let the origin on the manifold $\mathcal{M}$ be ${(\mr{\mbf{R}}, \mr{\sbf{\omega}}) := (\mbf{I}, \mbf{0})}$.
The system state can be calculated using the right state action and the origin as ${(\mbf{R}, \sbf{\omega}) = \phi((\mbf{Q},\mbf{q}), (\mathring{\mbf{R}}, \mathring{\sbf{\omega}}))}$.

\subsection{Equivariant Condition and Right Input Action}
\label{subsec:equivariant_condition_and_right_input_action}

We now define an input transformation ${\psi:G \times \mathbb{L}\to \mathbb{L}}$ which encodes the equivariance of the system. To achieve this, the equivariance condition 
\begin{equation}
\begin{split}
    D \phi_{(\mbf{Q},\mbf{q})}(\mbf{R}&,\sbf{\omega})[\bar{f}((\mbf{R},\sbf{\omega}),(\mbf{u},\mbf{a},\mbf{v},\mbf{w}))] \\
    &= \bar{f}(\phi_{(\mbf{Q},\mbf{q})}(\mbf{R},\sbf{\omega}),\psi_{(\mbf{Q},\mbf{q})}(\mbf{u},\mbf{a},\mbf{v},\mbf{w}))
\end{split}
    \label{eq:equivariance_condition_extended_system}
\end{equation}
must be satisfied. The left-hand side of \eqref{eq:equivariance_condition_extended_system} is given by
\begin{equation}
    \bigl(\mbf{R}(\mbf{u} - \sbf{\omega} + \mbf{v})^{\wedge}\mbf{Q},\, \mbf{Q}^\T (\sbf{\omega} + \mbf{w})\times \mbf{Q}^\T\mbf{u} + \mbf{Q}^\T \mbf{a}\bigr)
\label{eq:equivariance_condition_extended_system_left}
\end{equation}
and the right-hand side is given by
\begin{equation}
\begin{split}
    &\Bigl( \mbf{R}\mbf{Q} \bigl(\psi_{(\mbf{Q}, \mbf{q})}(\mbf{u}) - \mbf{Q}^\T(\sbf{\omega} - \mbf{q}) + \psi_{(\mbf{Q}, \mbf{q})}(\mbf{v})\bigr)^{\wedge} ,\\
    &\bigl(\mbf{Q}^\T(\sbf{\omega} - \mbf{q}) + \psi_{(\mbf{Q}, \mbf{q})}(\mbf{w})\bigr)\times \psi_{(\mbf{Q}, \mbf{q})}(\mbf{u}) \\
    &\hspace{50mm}+ \psi_{(\mbf{Q}, \mbf{q})}(\mbf{a})\Bigr).
\end{split}
    \label{eq:equivariance_condition_extended_system_right}
\end{equation}
From \eqref{eq:equivariance_condition_extended_system_left} and \eqref{eq:equivariance_condition_extended_system_right}, we conclude that the right action
\begin{equation}
\begin{split}
    \psi\bigl((\mbf{Q},\mbf{q})&, (\mbf{u}, \mbf{a}, \mbf{v}, \mbf{w})\bigr)\\ 
     &= \bigl(\mbf{Q}^\T \mbf{u}, \mbf{Q}^\T \mbf{a}, \mbf{Q}^\T (\mbf{v}- \mbf{q}), \mbf{Q}^\T (\mbf{w} + \mbf{q})\bigr)
\end{split}
    \label{eq:psi_input_group_action}
\end{equation}
satisfies the equivariance condition \eqref{eq:equivariance_condition_extended_system}.

\subsection{Equivariant Output and Right Output Action}
\label{subsec:equivariant_output_and_right_output_action}

As introduced in \ref{subsec:measurement_model}, the output of the system is composed of two non-collinear vectors, ${\mbf{y} = (\mbf{d}_1,\,\mbf{d}_2)}$. We define the right output action ${\rho:G\times \mathcal{Y}\to \mathcal{Y}}$ as
\begin{equation}
    \rho_{(\mbf{Q},\mbf{q})}(\mbf{d}_1, \mbf{d}_2) = \bigl(\mbf{Q}^\T\mbf{d}_1,\, \mbf{Q}^\T\mbf{d}_2\bigr)\,.
    \label{eq:right_output_action}
\end{equation}
Since ${\rho_{(\mbf{Q},\mbf{q})}(\mbf{d}_1, \mbf{d}_2) = h\bigl(\phi_{(\mbf{Q},\mbf{q})}(\mbf{R}, \sbf{\omega})\bigr)}$, we conclude that the output ${\mbf{y} = (\mbf{d}_1,\,\mbf{d}_2)}$ is equivariant \cite{mahony_observer_2022}.

\subsection{Equivariant System Lift}
\label{subsec:equivariant_system_lift}

To design the EqF, we first need to deduce an equivariant lift of the system. The lift ${\Lambda : \mathcal{M} \times \mathbb{L} \to \mathfrak{g}}$ is a map that will be used to lift the dynamics of the system from the state manifold, with inputs on the input manifold, to the Lie-algebra of the symmetry group \cite{mahony_observer_2022, mahony_equivariant_2021}. 

We propose the equivariant lift of the system
\begin{equation}
\begin{split}
    \Lambda((\mbf{R},& \sbf{\omega}), ( \mbf{u},\mbf{a},\mbf{v},\mbf{w}))\\
    &= ((\mbf{u} - \sbf{\omega} + \mbf{v})^{\wedge},\, -\mbf{a} + \mbf{u}\times\mbf{w} + \sbf{\omega}\times\mbf{v}) \,.   
\end{split}
    \label{eq:equivariant_lift}
\end{equation}
To be an equivariant lift of the system, ${\Lambda = (\Lambda_{\mbf{Q}},\Lambda_{\mbf{q}})}$ must satisfy the conditions
\begin{equation}
\begin{split}
    &D_{{(\mbf{Q}, \mbf{q})|_{(\mbf{I}, \mbf{0})}}}\phi_{(\mbf{R}, \sbf{\omega})}(\mbf{Q},\mbf{q})[(\Lambda_\mbf{Q},\Lambda_\mbf{q})] \\
    &\hspace{30mm}=\bar{f}((\mbf{R}, \sbf{\omega}), (\mbf{u},\mbf{a},\mbf{v},\mbf{w}))\,,
\end{split}
    \label{eq:equivariant_lift_condition_1}
\end{equation}
\begin{equation}
\begin{split}
    \text{Ad}&_{(\mbf{Q}, \mbf{q})^{-1}} (\Lambda_\mbf{Q},\Lambda_\mbf{q}) \\
    &\hspace{5mm}=\Lambda\bigl(\phi_{(\mbf{Q}, \mbf{q})}(\mbf{R}, \sbf{\omega}), \psi_{(\mbf{Q}, \mbf{q})}((\mbf{u}, \mbf{a}, \mbf{v}, \mbf{w}))\bigr),
\end{split}
    \label{eq:equivariant_lift_condition_2}
\end{equation}
according to \cite{mahony_equivariant_2020}. The proof that the proposed lift satisfies both conditions can be found in the Appendix.

We define $(\mbf{Q},\mbf{q})$ as the true lifted system state and $(\hat{\mbf{Q}},\hat{\mbf{q}})$ as the filter estimate of the state. 
The lifted system dynamics on the group are described by ${(\dot{\mbf{Q}}, \dot{\mbf{q}})=DL_{(\mbf{Q},\mbf{q})}(\Lambda_\mbf{Q},\Lambda_\mbf{q})}$\;\cite{mahony_equivariant_2021}. For the system at hand, following \eqref{eq:left_group_multiplication_differential}, this is expressed as  
\begin{equation}
\begin{split}
    \dot{\mbf{Q}} &= \mbf{Q}\, \Lambda_{\mbf{Q}}\bigl(\phi_{(\mr{\mbf{R}},\mr{\sbf{\omega}})}(\mbf{Q}, \mbf{q}), (\mbf{u}, \mbf{a}, \mbf{v}, \mbf{w})\bigr) \,,\\ 
    \dot{\mbf{q}} &= \mbf{Q}\,\Lambda_{\mbf{q}}\bigl(\phi_{(\mr{\mbf{R}},\mr{\sbf{\omega}})}(\mbf{Q}, \mbf{q}), (\mbf{u}, \mbf{a}, \mbf{v}, \mbf{w})\bigr)\,. 
    \label{eq:equivariant_lifted_system}
\end{split}
\end{equation}
The filter dynamics are expressed by 
\begin{equation}
    {(\dot{\hat{\mbf{Q}}}, \dot{\hat{\mbf{q}}})=DL_{(\hat{\mbf{Q}},\hat{\mbf{q}})}(\Lambda_\mbf{Q},\Lambda_\mbf{q}) + DR_{(\hat{\mbf{Q}},\hat{\mbf{q}})}(\Delta_\mbf{Q}, \delta_\mbf{q})}\,,
    \label{eq:equivariant_filter_dynamics_general}
\end{equation}
where the lifted system \eqref{eq:equivariant_lifted_system} is used as the internal model for the filter dynamics and ${\Delta =(\Delta_\mbf{Q}, \delta_\mbf{q}) \in \mathfrak{g}}$ are the correction terms to be designed. For this system, according to \eqref{eq:right_group_multiplication_differential}, the filter dynamics are 
\begin{equation}
\begin{split}
    \dot{\hat{\mbf{Q}}} &= \hat{\mbf{Q}}\, \Lambda_{\mbf{Q}}\bigl(\phi_{(\mr{\mbf{R}},\mr{\sbf{\omega}})}(\hat{\mbf{Q}}, \hat{\mbf{q}}), (\mbf{u}, \mbf{a}, \mbf{v}, \mbf{w})\bigr) + \Delta_\mbf{Q} \hat{\mbf{Q}}\,,\\ 
    \dot{\hat{\mbf{q}}} &= \hat{\mbf{Q}}\,\Lambda_{\mbf{q}}\bigl(\phi_{(\mr{\mbf{R}},\mr{\sbf{\omega}})}(\hat{\mbf{Q}}, \hat{\mbf{q}}), (\mbf{u}, \mbf{a}, \mbf{v}, \mbf{w})\bigr) \\
    & \hspace{47mm} + \Delta_\mbf{Q}\hat{\mbf{q}} + \delta_\mbf{q}\,. 
    \label{eq:equivariant_filter_dynamics_specific}
\end{split}
\end{equation}

For the scenario of interest, where the target's angular velocity, expressed in the target frame, is constant, the filter equations above reduce to
\begin{equation}
\begin{split}
    \dot{\hat{\mbf{Q}}} &= \hat{\mbf{Q}}(\mbf{u})^{\wedge} + (\hat{\mbf{q}})^{\wedge}\hat{\mbf{Q}}  + \Delta_\mbf{Q} \hat{\mbf{Q}}\,,\\
    \dot{\hat{\mbf{q}}} &= \Delta_\mbf{Q}\hat{\mbf{q}} + \delta_\mbf{q}\,,
    \label{eq:equivariant_filter_dynamics_reduced}
\end{split}
\end{equation}
where we have set the virtual inputs $\mbf{v} = \mbf{w} = \mbf{0}$, as previously explained.

%% file: sections/04_filter_design.tex
\section{EQUIVARIANT FILTER DESIGN} \label{sec:filter_design}

In this section, we derive the Equivariant Filter, as per \cite{van_goor_equivariant_2023}. We analyze the system errors and derive the linearized error dynamics. We also define and linearize the output residual, characterize the Riccati state matrix and derive the correction terms. Lastly, we study the convergence properties of the filter.

\subsection{System Error}
\label{subsec:system_error}
To derive the EqF, we need to characterize the system error. We will define the error on the group, on the manifold, and on the tangent space at the origin.

First, the system error on the group ${\mbf{E} = (\mbf{E}_{\mbf{Q}},\mbf{E}_{\mbf{q}})}$ is defined as
\begin{equation}
\begin{split}
    \mbf{E} &= (\mbf{E}_{\mbf{Q}},\mbf{E}_{\mbf{q}}) = (\mbf{Q},\mbf{q})\cdot(\hat{\mbf{Q}},\hat{\mbf{q}})^{-1}\\
    &= (\mbf{Q}\hat{\mbf{Q}}^\T, - \mbf{Q}\hat{\mbf{Q}}^\T \hat{\mbf{q}} + \mbf{q} )\,.
\end{split}
\label{eq:group_error}
\end{equation}

Building on \eqref{eq:group_error}, the state error on the manifold ${\mbf{e} = (\mbf{e}_{\mbf{R}}, \mbf{e}_{\sbf{\omega}}) \in \mathcal{M}}$ is obtained via the state action $\phi$, acting on the origin, according to 
\begin{equation}
\begin{split}
    \mbf{e} = (\mbf{e}_{\mbf{R}}, \mbf{e}_{\sbf{\omega}}) 
    = \phi_{(\mr{\mbf{R}}, \mr{\sbf{\omega}})}(\mbf{E}_{\mbf{Q}},\mbf{E}_{\mbf{q}})
    = (\mbf{E}_{\mbf{Q}},-\mbf{E}_{\mbf{Q}}^\T\mbf{E}_{\mbf{q}})\,.
\end{split}
\label{eq:manifold_state_error}
\end{equation}

Lastly, to define an error on the tangent space at the origin, we begin by fixing a local coordinate chart ${\vartheta:\mathcal{N}_{(\mr{\mbf{R}}, \mr{\sbf{\omega}})} \to \mathcal{T}_{(\mr{\mbf{R}}, \mr{\sbf{\omega}})}\mathcal{M}}$, where ${\mathcal{N}_{(\mr{\mbf{R}}, \mr{\sbf{\omega}})}\subset\mathcal{M}}$ is a neighborhood of the origin $(\mr{\mbf{R}}, \mr{\sbf{\omega}})$ and ${\mathcal{T}_{(\mr{\mbf{R}}, \mr{\sbf{\omega}})}\mathcal{M}\equiv \mathfrak{so}(3)\times \mathbb{R}^3}$ is the tangent space at the origin. Since $\mathbb{R}^3$ is the underlying space of $\mathfrak{so}(3)$, we will modify the coordinate chart so that ${\vartheta:\mathcal{N}_{(\mr{\mbf{R}}, \mr{\sbf{\omega}})} \to \mathbb{R}^3\times\mathbb{R}^3}$ instead, which will allow us to perform calculations in vector form. As such, let ${\sbf{\varepsilon} = (\sbf{\varepsilon}_{\mbf{R}},\sbf{\varepsilon}_{\sbf{\omega}}) \in \mathbb{R}^3 \times \mathbb{R}^3}$ be the local coordinates of the state error, given by
\begin{equation}
    \sbf{\varepsilon} = (\sbf{\varepsilon}_{\mbf{R}},\sbf{\varepsilon}_{\sbf{\omega}}) = \vartheta(\mbf{e}) = \bigl(\text{log}(\mbf{e}_{\mbf{R}})^{\vee}, {\mbf{e}_{\sbf{\omega}}}\bigr)\,.
    \label{eq:local_coordinates_error}
\end{equation}
We note that $\text{log}(\cdot)$ is chosen to be the principal branch of the matrix logarithm, as the logarithm is multi-valued. When ${\mathbf{e}_\mathbf{R}}$ is a rotation matrix corresponding to a \SI{180}{\degree} rotation about any axis, two of its eigenvalues are real and negative ($-1$). Thus, the matrix logarithm in this case is purely imaginary and the local error $\sbf{\varepsilon}_{\mbf{R}}$ is undefined.

For the derivations that follow, we will require ${\vartheta^{-1}:\mathbb{R}^3\times\mathbb{R}^3\to\mathcal{N}_{(\mr{\mbf{R}}, \mr{\sbf{\omega}})}}$, that is, the inverse of the local coordinate chart, which maps the local coordinates of the state error back to the global state error. The inverse map is
\begin{equation}
    \mbf{e} = \vartheta^{-1}(\sbf{\varepsilon}) = \bigl( \text{exp}({\sbf{\varepsilon}_{\mbf{R}}}^{\wedge}), {\sbf{\varepsilon}_{\sbf{\omega}}} \bigr)\,.
\end{equation}
We will also use a vector formulation of the local coordinates of the state
error ${\sbf{\varepsilon} = [\sbf{\varepsilon}_{\mbf{R}}\; \sbf{\varepsilon}_{\sbf{\omega}}]^\T\in \mathbb{R}^{6}}$ in the following calculations. 

\subsection{Error Dynamics}
\label{subsec:error_dynamics}
To design the EqF, we need to characterize the error dynamics of the system. We begin by calculating the time derivative of the error on the state manifold $\mbf{e}$, expressed as
\begin{equation}
        \dot{\mbf{e}} = D\phi_{\mbf{e}}\bigl(\Lambda(\mbf{e},\mr{\mbf{u}}) - \Lambda((\mr{\mbf{R}},\mr{\sbf{\omega}}),\mr{\mbf{u}})\bigr) - D\phi_{\mbf{e}}\Delta\,,
\label{eq:manifold_state_error_dynamics}
\end{equation}
where ${\mr{\mbf{u}} = \psi_{(\hat{\mbf{Q}},\hat{\mbf{q}})^{-1}}(\mbf{u},\mbf{a},\mbf{v},\mbf{w})}$. Having the dynamics of the error on the manifold and assuming that ${\mbf{e}(t)\in\mathcal{N}_{(\mr{\mbf{R}}, \mr{\sbf{\omega}})}}$ for all time, we make use of the fact that ${\sbf{\varepsilon} = \vartheta(\mbf{e})}$ to derive $\dot{\sbf{\varepsilon}}$, which is given by
\begin{equation}
\begin{split}
    &\dot{\sbf{\varepsilon}} = D\vartheta \cdot D\phi_{\mbf{e}}\bigl(\Lambda(\mbf{e},\mr{\mbf{u}}) - \Lambda((\mr{\mbf{R}},\mr{\sbf{\omega}}),\mr{\mbf{u}})\bigr) \\
    &\hspace{45mm}- D\vartheta \cdot D\phi_{\mbf{e}}\Delta\,.
\end{split}
\end{equation}
Now, we must linearize the error dynamics of $\sbf{\varepsilon}$ at zero. The linearized observer dynamics are given by 
\begin{equation}
\begin{split}
    \dot{\sbf{\varepsilon}} \approx &\: D_{\mbf{e}_{|(\mr{\mbf{R}}, \mr{\sbf{\omega}})}} \vartheta(\mbf{e}) \cdot D_{\mbf{E}_{|(\mbf{I}, \mbf{0})}}\phi_{(\mr{\mbf{R}}, \mr{\sbf{\omega}})}(\mbf{E}) \\
    &\cdot D_{\mbf{e}_{|(\mr{\mbf{R}}, \mr{\sbf{\omega}})}} \Lambda(\mbf{e}, \mr{\mbf{u}}) \cdot D_{\sbf{\varepsilon}_{|\mbf{0}}}\vartheta^{-1}(\sbf{\varepsilon})[\sbf{\varepsilon}] \\
    &- D_{\mbf{e}_{|(\mr{\mbf{R}}, \mr{\sbf{\omega}})}} \vartheta(\mbf{e}) \cdot D_{\mbf{E}_{|(\mbf{I}, \mbf{0})}}\phi_{(\mr{\mbf{R}}, \mr{\sbf{\omega}})}(\mbf{E})[\Delta]\,.
\end{split}
    \label{eq:preobserver_dynamics_linearization}
\end{equation} 
The full derivation of the expression \eqref{eq:preobserver_dynamics_linearization} can be found in the supplementary material. The final result, i.e., the linearized observer dynamics, is given by
\begin{equation}
    \dot{\sbf{\varepsilon}} \approx (-\sbf{\varepsilon}_{\sbf{\omega}}, (\hat{\mbf{q}})^{\wedge}\sbf{\varepsilon}_{\sbf{\omega}}) - D\vartheta(\mbf{e}) \cdot D\phi_{(\mr{\mbf{R}}, \mr{\sbf{\omega}})}(\mbf{E})[\Delta]\,.\\
\label{eq:local_error_derivative}    
\end{equation}
Using the vector notation for $\sbf{\varepsilon}$, we can write the pre-observer dynamics as ${\dot{\sbf{\varepsilon}} \approx \mr{\mbf{A}}_t\sbf{\varepsilon}}$, with the matrix ${\mr{\mbf{A}}_t}$ being
\begin{equation}
    \mr{\mbf{A}}_t = \begin{bmatrix}
        \mbf{0} & -\mbf{I}\\
        \mbf{0} & (\hat{\mbf{q}})^{\wedge}
    \end{bmatrix}\,.
\end{equation}

\subsection{Output Residual}
\label{subsec:output_residual}

The output of the system ${\mbf{y} = h(\mbf{R},\sbf{\omega})}$, as defined in the measurement model \eqref{eq:system_output_y}, can be rewritten as
\begin{equation}
    \mbf{y} = h(\mbf{R},\sbf{\omega}) = h\bigl(\phi_{(\hat{\mbf{Q}},\hat{\mbf{q}})}(\vartheta^{-1}(\sbf{\varepsilon}))\bigr)\,.
      \label{eq:output_function_1}
\end{equation}
Substituting $\sbf{\varepsilon} = \mbf{0}$ in \eqref{eq:output_function_1} yields
\begin{equation}
    h\bigl(\phi_{(\hat{\mbf{Q}},\hat{\mbf{q}})}(\vartheta^{-1}(\mbf{0}))\bigr) = h(\hat{\mbf{R}},\hat{\sbf{\omega}}) = \hat{\mbf{y}} \,.
    \label{eq:output_function_2}
\end{equation}
Similar to error-state Kalman filters, in the Equivariant Filter formulation, the output ${\mbf{y} = \mbf{y}(\sbf{\varepsilon})}$ is treated as a function of the error \eqref{eq:output_function_1}, while the predicted output ${\hat{\mbf{y}} = h(\hat{\mbf{R}},\hat{\sbf{\omega}})}$ is treated as an independent signal. The output residual is defined as
\begin{equation}
    \Tilde{\mbf{y}} = \mbf{y}(\sbf{\varepsilon}) - \hat{\mbf{y}}\,.
    \label{eq:output_residual}
\end{equation}
Considering $\Tilde{\mbf{y}}$ as a function of ${\sbf{\varepsilon}}$ and linearizing around ${\sbf{\varepsilon} = \mbf{0}}$, gives
\begin{equation}
\begin{split} 
    &\Tilde{\mbf{y}} \approx D_{(\mbf{R}, \sbf{\omega})_{|(\hat{\mbf{R}}, \hat{\sbf{\omega}})}}h(\mbf{R}, \sbf{\omega}) \,\cdot \\ 
    &\hspace{20mm}D_{\mbf{e}_{|(\mr{\mbf{R}}, \mr{\sbf{\omega})}}}\phi_{(\hat{\mbf{Q}}, \hat{\mbf{q}})}(\mbf{e}) \cdot
    D_{\sbf{\varepsilon}_{|\mbf{0}}}\vartheta^{-1}(\sbf{\varepsilon}) [\sbf{\varepsilon}] \,.
\end{split}
\label{eq:output_residual_linearization}
\end{equation}
The full derivation of the expression \eqref{eq:output_residual_linearization} is provided in the supplementary material. The final result is
\begin{equation}
\Tilde{\mbf{y}} = \Bigl(\hat{\mbf{Q}}^\T\bigl(\mr{\mbf{d}}_1\bigr)^{\wedge}\,\sbf{\varepsilon}_{\mbf{R}}, \hat{\mbf{Q}}^\T\bigl(\mr{\mbf{d}}_2\bigr)^{\wedge}\,\sbf{\varepsilon}_{\mbf{R}}\Bigr)\,.
\label{eq:output_residual_linearized}
\end{equation}
Using the vector notation for $\sbf{\varepsilon}$, we can write ${\Tilde{\mbf{y}} = \mbf{C}_t\sbf{\varepsilon}}$, with the matrix ${\mbf{C}_t}$ being
\begin{equation}
    \mbf{C}_t = 
    \begin{bmatrix}
    \hat{\mbf{Q}}^\T\bigl(\mr{\mbf{d}}_1\bigr)^{\wedge} & \mbf{0}\\
    \hat{\mbf{Q}}^\T\bigl(\mr{\mbf{d}}_2\bigr)^{\wedge} & \mbf{0}\\
    \end{bmatrix}\,.
\end{equation}

Since the output of this system is equivariant, the linearization of the output residual \eqref{eq:output_residual} can be refined to obtain an ${\mathcal{O}(\|\sbf{\varepsilon}\|^3)}$ error \cite{van_goor_equivariant_2023}. This improved linearization is given by
\begin{equation}
\begin{split}
    &\Tilde{\mbf{y}}  = \frac{1}{2}\Bigl(D_{(\mbf{E}_{\mbf{Q}},\mbf{E}_{\mbf{q}})_{|(\mbf{I},\mbf{0})}}\rho((\mbf{E}_{\mbf{Q}},\mbf{E}_{\mbf{q}}), \mbf{y}) \\
    &\hspace{4mm}+D_{(\mbf{E}_{\mbf{Q}},\mbf{E}_{\mbf{q}})_{|(\mbf{I},\mbf{0})}}\rho((\mbf{E}_{\mbf{Q}},\mbf{E}_{\mbf{q}}), \hat{\mbf{y}})\Bigr) \cdot  \text{Ad}_{(\hat{\mbf{Q}},\hat{\mbf{q}})^{-1}}[\sbf{\varepsilon}^{\wedge}].
\end{split}
    \label{eq:output_residual_linearization_improved}
\end{equation}
The full derivation of the expression \eqref{eq:output_residual_linearization_improved} is shown in the supplementary material. The linearized output residual is
\begin{equation}
    \Tilde{\mbf{y}} = \frac{1}{2}\bigl((\mbf{d}_1 + \hat{\mbf{d}}_1)^{\wedge}\,\hat{\mbf{Q}}^\T \sbf{\varepsilon}_{\mbf{R}},\, (\mbf{d}_2 + \hat{\mbf{d}}_2)^{\wedge}\,\hat{\mbf{Q}}^\T \sbf{\varepsilon}_{\mbf{R}} \bigr)\,.
\end{equation}
Again, using the vector notation for $\sbf{\varepsilon}$, we can write ${\Tilde{\mbf{y}} = \mbf{C}_t^*\sbf{\varepsilon}}$, with the matrix ${\mbf{C}_t^*}$ being
\begin{equation}
    {\mbf{C}_t^*} = \frac{1}{2} 
    \begin{bmatrix}
    \bigl(\mbf{d}_1 + \hat{\mbf{d}}_1\bigr)^{\wedge} \hat{\mbf{Q}}^\T & \mbf{0}\\
    \bigl(\mbf{d}_2 + \hat{\mbf{d}}_2\bigr)^{\wedge} \hat{\mbf{Q}}^\T& \mbf{0}\\
    \end{bmatrix}\,.
\end{equation}

\subsection{Riccati State}
\label{subsec:riccati_state}
The correction terms ${(\Delta_{\mbf{Q}}, \delta_{\mbf{q}})}$ and, more precisely, the direction of the Fréchet derivative used in their calculation depend on the Riccati state ${\sbf{\Sigma}\in \mathbb{S}_{+}^6}$, ${\mathbb{S}_{+}^6}$ the set of positive definite ${6\times 6}$ matrices. This state evolves according to
\begin{equation}
\begin{split}
    &\dot{\sbf{\Sigma}} =  \mr{\mbf{A}}_t \sbf{\Sigma} + \sbf{\Sigma} \mr{\mbf{A}}_t^\T + \mbf{M}_t - \sbf{\Sigma} \mbf{C}_t^\T \mbf{N}_t^{-1} \mbf{C}_t \sbf{\Sigma}\,,\\
    &\sbf{\Sigma}(0) = \sbf{\Sigma}_0\,,
    \label{eq:riccati_state_equation}
\end{split}
\end{equation}
where ${\sbf{\Sigma}_0 \in\mathbb{S}_{+}^6}$ is the initial value of $\sbf{\Sigma}$, which reflects the uncertainty in the initial estimate in the local coordinates. The matrices ${\mbf{M}_t \in\mathbb{S}_{+}^6}$ and ${\mbf{N}_t \in\mathbb{S}_{+}^{6}}$ represent the state gain and output gain matrices, respectively. The matrix $\mbf{M}_t$  is associated with the process noise, while the matrix $\mbf{N}_t$ is associated with the measurement noise.

The Riccati state matrix ${\sbf{\Sigma}}$ can be written in block form as 
\begin{equation}
    \sbf{\Sigma} = \begin{bmatrix}
        \sbf{\Sigma}_{\mbf{R}} & \sbf{\Sigma}_{\mbf{R},\sbf{\omega}}\\
        \sbf{\Sigma}_{\mbf{R},\sbf{\omega}}^\T & \sbf{\Sigma}_{\sbf{\omega}}\\
    \end{bmatrix},
    \label{eq:riccati_state_block}
\end{equation}
with ${\sbf{\Sigma}_{\mbf{R}}, \sbf{\Sigma}_{\sbf{\omega}} \in \mathbb{S}_{+}^3}$ and ${\sbf{\Sigma}_{\mbf{R},\sbf{\omega}}\in \mathbb{R}^{3\times 3}}$.

\subsection{Correction Terms}
\label{subsec:correction_terms}

Let us now more thoroughly analyze the correction terms ${(\Delta_{\mbf{Q}}, \delta_{\mbf{q}})}$, which were introduced in \eqref{eq:equivariant_filter_dynamics_general}. Consider a direction ${\sbf{\gamma} = [\sbf{\gamma}_1 \,\sbf{\gamma}_2]^\T\in\mathbb{R}^{6}}$, with ${\sbf{\gamma}_1,\,\sbf{\gamma}_2\in\mathbb{R}^3}$, given by 
\begin{equation}
    \sbf{\gamma} = \sbf{\Sigma} \mbf{C}_t^\T \mbf{N}_t^{-1}\bigl(\mbf{y}-\hat{\mbf{y}}\bigr)\,, 
    \label{eq:correction_term_delta_direction}
\end{equation}
where we have used the vector form of $\mbf{y}$ and $\hat{\mbf{y}}$ and ${\sbf{\Sigma} \mbf{C}_t^\T \mbf{N}_t^{-1}}$ acts as the Kalman gain. For this system, since the output is equivariant, we can substitute the matrix ${\mbf{C}_t^*}$ for the matrix ${\mbf{C}_t}$ in \eqref{eq:correction_term_delta_direction}. The correction terms $(\Delta_{\mbf{Q}},\delta_{\mbf{q}})$ are then calculated according to
\begin{equation}
\begin{split}
    (\Delta_{\mbf{Q}},\delta_{\mbf{q}}) &= D_{\mbf{E}_{|(\mbf{I}, \mbf{0})}} \phi_{(\mr{\mbf{R}},\mr{\sbf{\omega}})}(\mbf{E})^\dagger \cdot D\vartheta^{-1} [\sbf{\gamma}]\\
    &= ({\sbf{\gamma}_1}^{\wedge},-{\sbf{\gamma}_2})\,,
    \label{eq:correction_term_delta}
\end{split}
\end{equation}
where $(\cdot)^\dagger$ denotes a right inverse.
The complete derivation of \eqref{eq:correction_term_delta} can be found in the supplementary material. Considering that ${\mbf{N}_t = \text{diag}([k_N^{-1}\;...\;\;k_N^{-1}])}$, with ${k_N \in\mathbb{R}_{+}}$ and that the Riccati state has the block form \eqref{eq:riccati_state_block}, the correction terms are
\begin{equation}
    \begin{split}
        \Delta_\mbf{Q} &= k_N \bigl( \sbf{\Sigma}_\mbf{R} \cdot (\hat{\mbf{Q}}^\T \mr{\mbf{d}}_1 \times \mr{\mbf{d}}_1 + \hat{\mbf{Q}}^\T \mr{\mbf{d}}_2 \times \mr{\mbf{d}}_2) \bigr)^{\wedge}\,,\\
        \delta_\mbf{q} &= -k_N \cdot \sbf{\Sigma}_{\mbf{R}\sbf{\omega}}^\T \cdot (\hat{\mbf{Q}}^\T \mr{\mbf{d}}_1 \times \mr{\mbf{d}}_1 + \hat{\mbf{Q}}^\T \mr{\mbf{d}}_2 \times \mr{\mbf{d}}_2)\,.
    \end{split}
\end{equation}
In this case, using either the matrix $\mbf{C}_t$ or $\mbf{C}_t^*$ yields the same correction terms.

\subsection{Convergence Analysis} \label{subsec:convergence_analysis}

Since $\mr{\mbf{A}}_t$ depends on $\hat{\mbf{q}}$, if ${\hat{\mbf{q}}}$ remains bounded for all time, then the pair ${(\mr{\mbf{A}}_t, \mbf{C}_t)}$ is uniformly completely observable\;\cite{bristeau_design_2010} and the second derivative of the error dynamics is bounded. Consequently, the Riccati state ${\sbf{\Sigma}(t)}$ is bounded above and below, meaning that \eqref{eq:riccati_state_equation} is well-defined for all time \cite{deylon_uniform_observability_2001}. 
Provided that the initial error ${\sbf{\varepsilon}(0)}$ is sufficiently small, these conditions guarantee that ${\sbf{\varepsilon} \to \mbf{0}}$ or, equivalently, that ${\mbf{e}\to(\mr{\mbf{R}},\mr{\sbf{\omega}})}$ and ${\mbf{E}\to(\mbf{I},\mbf{0})}$ \cite{krener_convergence_2003}. We can define the Lyapunov function
\begin{equation}
    V = \sbf{\varepsilon}^\T \sbf{\Sigma}^{-1}\sbf{\varepsilon}\,.
    \label{eq:lyapunov_function}
\end{equation}
Combining \eqref{eq:local_error_derivative}, \eqref{eq:correction_term_delta_direction} and \eqref{eq:correction_term_delta}, it follows that
\begin{equation}
    \dot{\sbf{\varepsilon}} \approx \mr{\mbf{A}}_t \sbf{\varepsilon} - \sbf{\Sigma}\mbf{C}^\T \mbf{N}_t^{-1}\mbf{C}_t \sbf{\varepsilon}\,.
\end{equation}
Using the above result and taking the time-derivative of the Lyapunov function, then
\begin{equation}
    \begin{split}
        \dot{V} &= \dot{\sbf{\varepsilon}}^\T \sbf{\Sigma}^{-1}\sbf{\varepsilon} - \sbf{\varepsilon}^\T \sbf{\Sigma}^{-1} \dot{\sbf{\Sigma}}\sbf{\Sigma}^{-1}\sbf{\varepsilon} + \sbf{\varepsilon}^\T \sbf{\Sigma}^{-1}\dot{\sbf{\varepsilon}} \\
        &= -\sbf{\varepsilon}^\T\bigl( \sbf{\Sigma}^{-1} \mbf{M}_t \sbf{\Sigma}^{-1} + \mbf{C}_t^\T \mbf{N}_t^{-1} \mbf{C}_t \bigr) \sbf{\varepsilon} \,,
        \end{split}
        \label{eq:lyapunov_function_derivative}
\end{equation}
from which we conclude that ${\dot{V} < 0}$, ${\sbf{\varepsilon}\neq\mbf{0}}$, and $\sbf{\varepsilon}$ converges to zero as ${t \to \infty}$.

%% file: sections/05_simulation_results.tex
\section{SIMULATION RESULTS} \label{sec:simulation_results}

To evaluate the performance of the Equivariant Filter derived in Section\;\ref{sec:filter_design}, we conduct numerical simulations using MATLAB\textsuperscript{\textregistered}. To ensure that the estimates remain in the group $G$, instead of using a solver to simulate the system, we discretize and implement the system and filter equations. 

\subsection{Simulation Setup}
\label{subsec:simuulation_setup}

\begin{figure}
    \centering
    \begin{minipage}{\columnwidth}
            \centering
            \begin{subfigure}[t]{\textwidth}
                \centering
                \includegraphics[width=\textwidth]{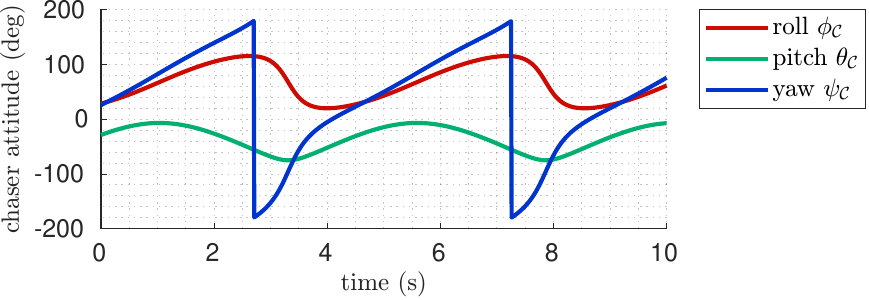} 
                \caption{Chaser attitude, $^\mathcal{I}_\mathcal{C}\mbf{R}$.}
                \label{fig:simulation_attitude_chaser}
            \end{subfigure}
            \begin{subfigure}[t]{\textwidth}
                \centering
                \includegraphics[width=\textwidth]{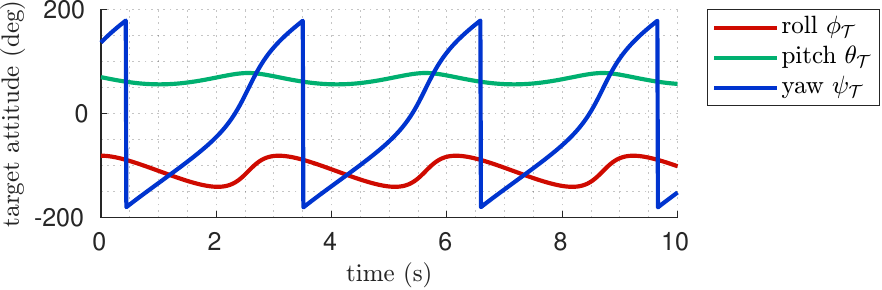} 
                \caption{Target attitude, $^\mathcal{I}_\mathcal{T}\mbf{R}$.}
                \label{fig:simulation_attitude_target}
            \end{subfigure}
    \end{minipage}
    \caption{Chaser and target attitudes, expressed in Euler angles.}
    \label{fig:simulation_attitudes}
\end{figure}

We initialize the chaser and target with random attitudes, which leads to a random initial relative attitude $\mbf{R}$. To recover the system of interest, we set the target's angular acceleration to zero, ${\mbf{a} = \mbf{0}}$, as well as the virtual inputs that were introduced to prove equivariance, ${\mbf{v} = \mbf{w} = \mbf{0}}$. Hence, we assume a random constant angular velocity $\sbf{\omega}_\mathcal{T}$ for the target. The chaser's angular velocity, input $\mbf{u}$, is also constant and is generated randomly. In Fig.\;\ref{fig:simulation_attitudes}, we display the chaser and target attitudes, with respect to the inertial frame, throughout the simulation, which resulted from the random choice of constant angular velocities. For visualization purposes, the attitudes are presented in Euler angles.

The initial filter estimates in the group are set to ${(\hat{\mbf{Q}},\hat{\mbf{q}}) = (\mbf{I}, \mbf{0}})$, with gain matrices ${\mbf{M}_t = \mbf{I}}$ and ${\mbf{N}_t = 0.1\,\mbf{I}}$. The initial value of the Riccati state is ${\sbf{\Sigma}_0 = \mbf{I}}$. The chosen reference vectors, constant on the target frame ${\{\mathcal{T}\}}$, are ${\mr{\mbf{d}}_1 = [1\;0\;0]^\T}$ and ${\mr{\mbf{d}}_2 = [0\;1\;0]^\T}$. As explained in\;\ref{subsec:measurement_model}, measurement noise is considered so that the measurement is a rotation of the ideal measurement by an angle $\theta$ about an axis $\mbf{n}$. In the simulations, the angle $\theta$ follows a normal distribution, specifically ${\theta \sim \mathpzc{N}(0, 0.1^2)}$, and the axis follows a uniform distribution on $\mathcal{S}^2$, ${\mbf{n}\sim \mathpzc{U}(\mathcal{S}^2)}$.

The Equivariant Filter is implemented in a prediction-update formulation. The system lifted onto the group, as described in \eqref{eq:equivariant_lifted_system}, replicated in the filter equations \eqref{eq:equivariant_filter_dynamics_specific}, is used in the prediction step, employing the input $\mbf{u}$ and the filter state ${(\hat{\mbf{Q}},\hat{\mbf{q}})}$ at that time. In the update step, the measurements ${(\mbf{d}_1,\mbf{d}_2)}$ are acquired and the correction terms ${(\Delta_\mbf{Q},\delta_\mbf{q})}$ are calculated and applied. The simulation was run with a time step of {\SI{0.01}{\second}}, with both prediction and update steps being calculated sequentially at each instant. 

\subsection{Equivariant Filter Performance}
\label{subsec:equivariant_filter_performance}

\begin{figure}
    \centering
    \begin{minipage}{\columnwidth}
            \centering
            \begin{subfigure}[t]{\textwidth}
                \centering
                \includegraphics[width=\textwidth]{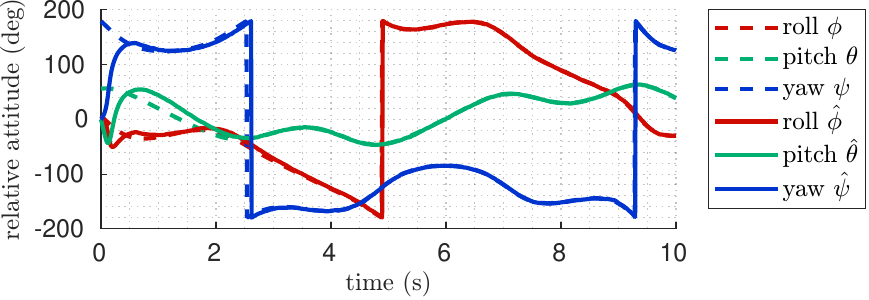} 
                \caption{Relative attitude, $\mbf{R}$, expressed in Euler angles.}
                \label{fig:relative_attitude}
            \end{subfigure}
            \begin{subfigure}[t]{\textwidth}
                \centering
                \includegraphics[width=\columnwidth]{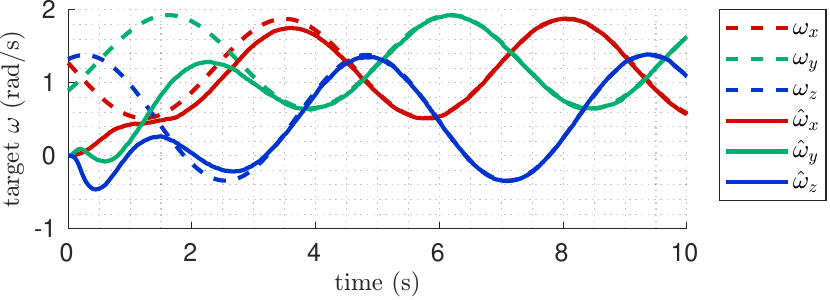}
                \caption{Target's angular velocity, $\sbf{\omega}$.}
                \label{fig:angular_velocity_target}
            \end{subfigure}
    \end{minipage}
    \caption{True relative attitude and target's angular velocity ${(\mbf{R},\sbf{\omega})}$ (dashed) and estimated ${(\hat{\mbf{R}},\hat{\sbf{\omega}})}$ (solid).}
    \label{fig:simulation_results}
\end{figure}

The estimates of the relative attitude and the target's angular velocity, expressed on the chaser frame, i.e., the states on the manifold $\mathcal{M}$, are displayed in Fig.\;\ref{fig:simulation_results}. The true values $(\mbf{R}, \sbf{\omega})$ are depicted in dashed lines, while the estimates $(\hat{\mbf{R}}, \hat{\sbf{\omega}})$ are depicted in solid lines. As before, for illustrative purposes, the relative attitude is represented in Euler angles. As expected, the filter estimates converge to the true values.

\begin{figure}
    \centering
    \begin{minipage}{\columnwidth}
       \centering
        \begin{subfigure}[t]{\textwidth}
            \centering
            \includegraphics[width=\textwidth]{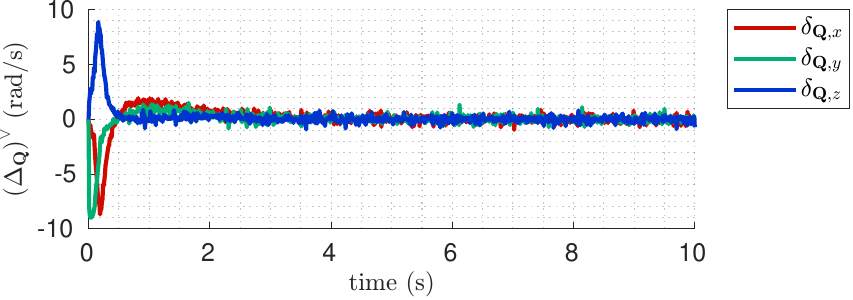} 
            \caption{Correction term $\delta_\mbf{Q}={\Delta_\mbf{Q}}^{\vee}$.}
            \label{fig:correction_terms_Q}
        \end{subfigure}
        \begin{subfigure}[t]{\textwidth}
            \centering
            \includegraphics[width=\textwidth]{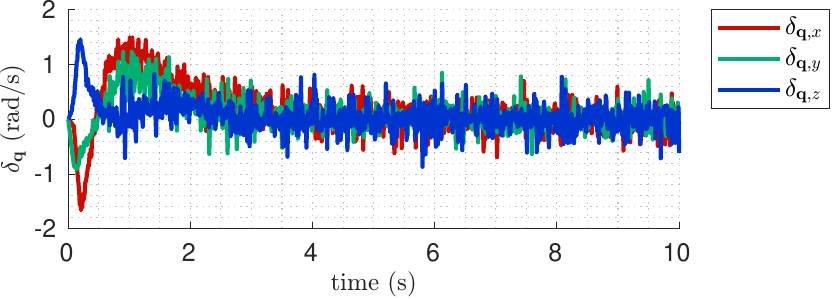}
            \caption{Correction term $\delta_\mbf{q}$.}
            \label{fig:correction_terms_q}
        \end{subfigure}
    \end{minipage}
    \caption{Filter correction terms.}
    \label{fig:correction_terms}
\end{figure}

The correction terms ${(\Delta_\mbf{Q},\delta_\mbf{q})}$ are presented in Fig.\;\ref{fig:correction_terms}. To aid in visualization, we map $\Delta_\mbf{Q}$ to $\mathbb{R}^3$, with ${\delta_\mbf{Q}={\Delta_\mbf{Q}}^{\vee}}$. We can observe that the noise in the measurements is reflected in the correction terms, which approximate zero as the estimates converge to the true values, making the filter more closely mimic the true system. 

\begin{figure}
    \centering
    \includegraphics[width=\columnwidth]{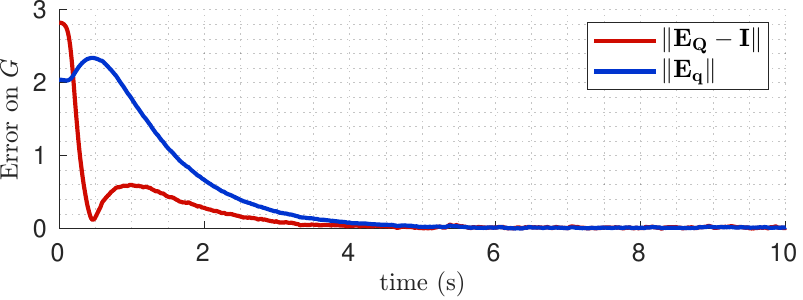}
    \caption{Norm of the error on the group (with the identity subtracted).}
    \label{fig:error_group}
\end{figure}

In Fig.\;\ref{fig:error_group}, we present the error norm on the group with the identity element subtracted, that is, ${\|\mbf{E}_{\mbf{Q}}-\mbf{I}\|}$ and ${\|\mbf{E}_{\mbf{q}}\|}$. Since the group error is expected to converge to the identity, subtracting the identity simplifies the interpretation of the norm, making it easier to assess the performance of the filter as it approaches zero. The error norm on the group converges to zero after approximately \SI{4}{\second}, which is consistent with the results observed in Fig.\;\ref{fig:simulation_results}. 
After converging, the mean of the relative attitude error norm is \num{0.019}, which corresponds to a mean error norm of \SI{0.572}{\degree}, \SI{0.437}{\degree}, and \SI{0.435}{\degree} for roll, pitch and yaw, respectively.
The mean of the target's angular velocity error norm is \SI{0.023}{\radian\per\second}, or \SI{1.318}{\degree\per\second}, that is, a relative error of \SI{1.11}{\percent} with respect to the true target's angular velocity norm.

\subsection{Monte Carlo Simulations}
\label{subsec:monte_carlo_simulations}
For a broader analysis of the filter in the simulation environment, we conduct Monte Carlo simulations, each run with a randomly generated initial relative attitude, and random constant angular velocities $\sbf{\omega}_\mathcal{T}$ and $\mbf{u}$. The measurement noise distributions are the same as before. We also keep the initial values for the Riccati state, ${\sbf{\Sigma}_0 = \mbf{I}}$, and state and output gain matrices, ${\mbf{M}_t = \mbf{I}}$ and ${\mbf{N}_t = 0.1\,\mbf{I}}$, for all simulations. 

\begin{figure}
    \centering
    \includegraphics[width=\columnwidth]{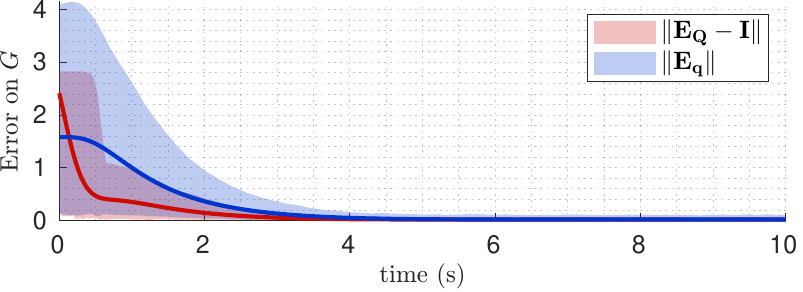}
    \caption{Monte Carlo simulations: Mean error (solid lines) and error range (shaded areas) on the group.}
    \label{fig:monte_carlo_error_group}
\end{figure}

For this purpose, we define a successful run as the filter converging in less than \SI{10}{\second} to an estimate which satisfies $\|\mbf{E}_{\mbf{Q}} - \mbf{I}\|<0.1$ and $\|\mbf{E}_{\mbf{q}}\|<\SI{0.1}{\radian\per\second}$. A total of \num{1000} Monte Carlo simulations were run, with \num{1} failure detected, which gives us a \SI{99}{\percent} confidence interval with a \SI{1}{\percent} margin of error.
The mean attitude error and the mean target's angular velocity error on the group, across all simulations, are depicted in Fig.\;\ref{fig:monte_carlo_error_group}. The shaded areas illustrate the range of values observed across all simulations. 
In general, throughout the simulations, at approximately \SI{4}{\second}, the filter has converged. After this time, the mean of the relative attitude error norm is \num{0.020} and the mean of the target's angular velocity error norm is \SI{0.024}{\radian\per\second}, or \SI{1.375}{\degree\per\second}.

\subsection{Comparison with an Extended Kalman Filter}
\label{subsec:comparison_ekf}

\begin{figure}
    \centering
    \begin{minipage}{\columnwidth}
            \centering
            \begin{subfigure}[t]{\textwidth}
                \centering
                \includegraphics[width=\textwidth]{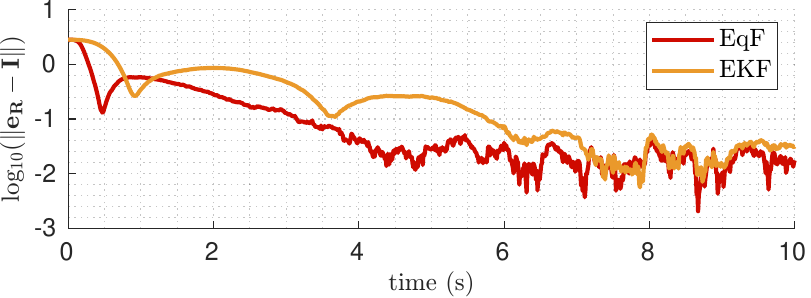} 
                \caption{Relative attitude error on the manifold.}
                \label{fig:error_R_eqf_vs_ekf_log}
            \end{subfigure}
            \begin{subfigure}[t]{\textwidth}
                \centering
                \includegraphics[width=\textwidth]{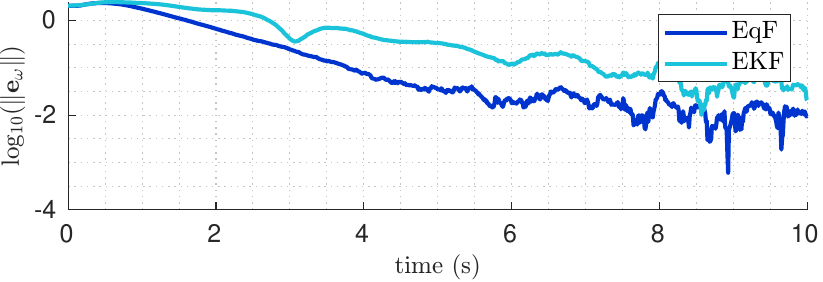} 
                \caption{Target's angular velocity error on the manifold.}
                \label{fig:error_omega_eqf_vs_ekf_log}
            \end{subfigure}
    \end{minipage}
    \caption{Comparison between the logarithm of the errors of the EqF and EKF on the manifold.}
    \label{fig:error_eqf_vs_ekf_log}
\end{figure}

Let us now compare the performance of the EqF with that of an EKF, under the same conditions as in the first scenario. In Fig.\;\ref{fig:error_eqf_vs_ekf_log}, we present the logarithm of the errors in the relative attitude and the target's angular velocity. It is evident that the error of the EqF decreases faster than the EKF error.

In Tables\;{\ref{tab:eqf_ekf_performance_metrics_relative_attitude}} and {\ref{tab:eqf_ekf_performance_metrics_angular_velocity}}, we present three metrics comparing the EqF and EKF performances in estimating the relative attitude and the target's angular velocity, respectively. Table\;{\ref{tab:eqf_ekf_performance_metrics_relative_attitude}} contains the mean error (in\;${\si{\degree}}$) for the relative roll, pitch and yaw angles, in the time interval between 4 and 10\,${\si{\second}}$, the minimum error (in\;${\si{\degree}}$) for the entire simulation, and the time (in\;${\si{\second}}$) that it takes for the filter to achieve a ${\SI{1}{\degree}}$ error in each of the angles. Table\;{\ref{tab:eqf_ekf_performance_metrics_angular_velocity}} similarly contains the mean error norm (in\;${\si{\radian\per\second}}$) for the target's angular velocity, in the time interval between 4 and 10\,${\si{\second}}$, the minimum error norm (in\;${\si{\radian\per\second}}$) for the entire simulation, and the time (in\;${\si{\second}}$) that it takes for the filter to achieve a ${\SI{0.1}{\radian\per\second}}$ error.
{
\sisetup{
  detect-all,
  round-mode=places,
  round-precision=2,
  separate-uncertainty=true,
  output-open-uncertainty = [,
  output-close-uncertainty = ],
}
\begin{table}
    \centering
    \caption{Comparison of EqF and EKF performance metrics: relative attitude estimation}
    \begin{tabular}{
        l  
        l  
        S[table-format=1.3]  
        S[scientific-notation=true, table-format=1.2e1]  
        S[table-format=1.2]  
        }
        \toprule
        \multicolumn{2}{c}{} &
        \multicolumn{1}{c}{\shortstack{Mean error (\si{\degree})\\in [4, 10]\,\si{\second}}} &
        \multicolumn{1}{c}{\shortstack{Minimum\\error (\si{\degree})}} &
        \multicolumn{1}{c}{\shortstack{Time to\\1\si{\degree} error (\si{\second})}} \\
        \midrule
        
        \multirow{3}{*}{EqF} 
          & Roll $\phi$   & 0.572 & 3.27e-4 & 0.44 \\
          & Pitch $\theta$& 0.437 & 8.31e-4 & 0.46 \\
          & Yaw $\psi$    & 0.435 & 2.54e-4 & 0.50 \\
        \midrule
        \multirow{3}{*}{EKF}
          & Roll $\phi$   & 3.815 & 1.986e-3 & 0.05 \\
          & Pitch $\theta$& 1.015 & 2.998e-3 & 0.92 \\
          & Yaw $\psi$    & 2.171 & 1.481e-2 & 0.84 \\
        \bottomrule
    \end{tabular}
    \label{tab:eqf_ekf_performance_metrics_relative_attitude}
\end{table}
}
{
\sisetup{
  detect-all,
  round-mode=places,
  round-precision=2,
  separate-uncertainty=true,
  output-open-uncertainty = [,
  output-close-uncertainty = ],
}
\begin{table}
    \centering
    \caption{Comparison of EqF and EKF performance metrics: target's angular velocity estimation}
    \begin{tabular}{l
                    S[table-format=1.3]
                    S[scientific-notation=true, table-format=1.2e1]
                    S[table-format=1.2]}
        \toprule
        {} & 
        {\shortstack{Mean error\\($\si{\radian\per\second}$)\\in [4, 10]\,s}} & 
        {\shortstack{Minimum\\error\\($\si{\radian\per\second}$)}} & 
        {\shortstack{Time ($\si{\second}$)\\to 0.1\,$\si{\radian\per\second}$\\ error}} \\
        \midrule
        EqF & 0.024 & 6.08134e-4 & 3.79 \\
        EKF & 0.159 & 1.00891e-2 & 7.13 \\
        \bottomrule
    \end{tabular}
    \label{tab:eqf_ekf_performance_metrics_angular_velocity}
\end{table}
}

Despite the advantage of a simpler derivation, the EKF formulation presents many drawbacks for the system at hand. Unlike the EqF, whose estimates evolve on the Lie group ${G=SE(3)}$, the estimates of the EKF evolve in ${\mathbb{R}^n}$, with ${n = 9+3 = 12}$, since we want to estimate the entries of a rotation matrix and a 3-dimensional vector. One could select another parametrization of the attitude, with fewer dimensions, though for consistency, we chose to keep the rotation matrix parametrization. By not working on the group, the estimate of the relative attitude will not remain a rotation matrix. Nonetheless, this can be overcome by projecting the estimate to ${\mathcal{SO}(3)}$ after each update. However, this is still less effective than performing the estimation directly on the group \cite{barrau_invariant_2017}. Furthermore, the convergence of the EKF is much more dependent on the state and output gain matrices than in the case of the EqF. As such, careful tuning of these gains is required under different conditions of the system. 

\subsection{Effect of Measurement Rate in Filter Performance}
\label{subsec:effect_of_measurement_rate_in_filter_performance}

Due to the safety specifications for spacecraft hardware, the processing capabilities and, more importantly for our purposes, the sensor acquisition rates are constrained \cite{eickhoff_onboard_2011}. Therefore, we analyze the effect that a low measurement rate has on the Equivariant Filter, using the same initial conditions and motion as in the previous scenario.

\begin{figure}
    \centering
    \begin{minipage}{\columnwidth}
            \centering
            \begin{subfigure}[t]{\textwidth}
                \centering
                \includegraphics[width=\textwidth]{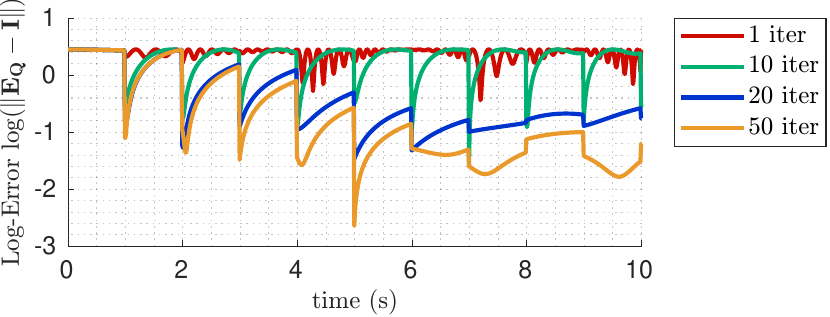} 
                \caption{Log-Error norm $\text{log}(\|\mbf{E}_{\mbf{Q}} - \mbf{I}\|)$ for measurement rate of \SI{1}{\hertz}.}
                \label{fig:multirate_1hz_error_Q}
            \end{subfigure}
            \begin{subfigure}[t]{\textwidth}
                \centering
                \includegraphics[width=\textwidth]{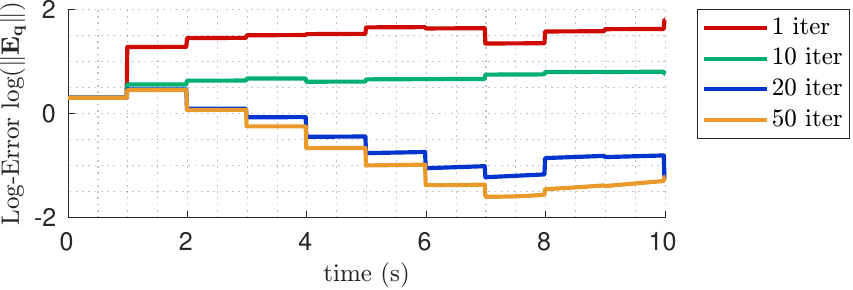} 
                \caption{Log-Error norm $\text{log}(\|\mbf{E}_{\mbf{q}}\|)$ for measurement rate of \SI{1}{\hertz}.}
                \label{fig:multirate_1hz_error_q}
            \end{subfigure}
    \end{minipage}
    \caption{Log-Error norm on the group (with the identity subtracted), for a measurement rate of \SI{1}{\hertz}, with the update step iterated over once (red), 10 (green), 20 (blue) and 50 times (yellow).}
    \label{fig:multirate_1hz_errors}
\end{figure}

\begin{figure}
    \centering
    \begin{minipage}{\columnwidth}
            \centering
            \begin{subfigure}[t]{\textwidth}
                \centering
                \includegraphics[width=\textwidth]{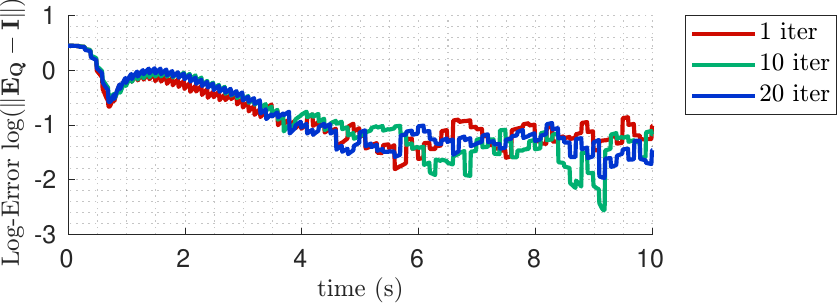} 
                \caption{Log-Error norm $\text{log}(\|\mbf{E}_{\mbf{Q}} - \mbf{I}\|)$ for measurement rate of \SI{30}{\hertz}.}
                \label{fig:multirate_30hz_error_Q}
            \end{subfigure}
            \begin{subfigure}[t]{\textwidth}
                \centering
                \includegraphics[width=\textwidth]{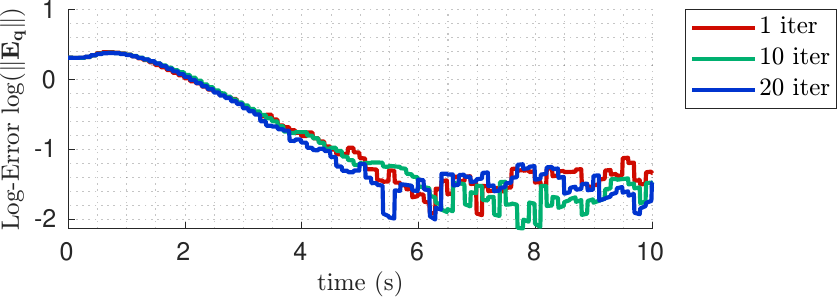} 
                \caption{Log-Error norm $\text{log}(\|\mbf{E}_{\mbf{q}}\|)$ for measurement rate of \SI{30}{\hertz}.}
                \label{fig:multirate_30hz_error_q}
            \end{subfigure}
    \end{minipage}
    \caption{Log-Error norm on the group (with the identity subtracted), for a measurement rate of \SI{30}{\hertz}, with the update step iterated over once (red), 10 (green), and 20 times (blue).}
    \label{fig:multirate_30hz_errors}
\end{figure}

We consider an acquisition frequency of \SI{1}{\hertz}, typical for cameras equipped on spacecraft \cite{colmenarejo_ground_2019}, which is the sensor we assume measures the vectors ${(\mbf{d}_1,\mbf{d}_2)}$. We also study the scenario of an acquisition rate of \SI{30}{\hertz}, a common frequency for commercial-grade cameras. These measurement acquisition rates mean that the filter's update step is performed at a lower frequency and fewer times than the prediction step, which in these simulations is calculated at a rate of \SI{100}{\hertz}, as before. The period considered for the prediction and update steps in the discrete-time implementation of the EqF is simply the inverse of the corresponding rate. In Figs.\;\ref{fig:multirate_1hz_errors} and \ref{fig:multirate_30hz_errors}, we present the log-error norm for the simulations with a measurement rate of \SI{1}{\hertz} and \SI{30}{\hertz}, respectively. 

As one could expect, having fewer measurements to correct the estimates can severely degrade the filter's performance, as observed in Fig.\;\ref{fig:multirate_1hz_errors}. To mitigate this effect, after each measurement, we can run the update step several times, instead of just once, using the same measurement and decreasing the update step's period. Here, we decrease it by a factor equal to the number of iterations performed. When the update step is executed, the correction terms ${(\Delta_\mbf{Q}, \delta_\mbf{q})}$, which depend on the measurements and the filter estimates, are calculated and the filter's states ${(\hat{\mbf{Q}},\hat{\mbf{q}})}$ are corrected. If we repeat this step, the estimates become more refined with each iteration, and the filter's performance improves. This effect is noticeable in the plots for simulations where the update step is iterated over once, ten, twenty, and fifty times. For a measurement rate of \SI{1}{\hertz}, with few iterations of the update step, the filter does not converge to the true value. However, if we increase the number of iterations, the error decreases as the filter is able to more closely estimate the relative attitude and the target's angular velocity. From Fig.\;\ref{fig:multirate_30hz_errors}, we conclude that a rate of \SI{30}{\hertz} is sufficiently fast for an accurate estimation of the motion and iterating the update step has practically no effect. 

Since the filter state dimension does not grow over time, the number of operations in each filter step and the respective computational complexity remain constant.
For a given system, the number of iterations would be constrained by the onboard processing power. Implemented in C-code and running on an Intel\textsuperscript{\textregistered} Core\texttrademark{} i7-10750H CPU, on average, the prediction step runs in \SI{9}{\micro\second} and the update step in \SI{73}{\micro\second}. 
Based on previous findings \cite{lourenco_vv4rtos_2023}, the expected runtime on a space-grade processor is approximately 140 to 211 times longer than on a commercial-grade processor. As such, at a measurement frequency of \SI{1}{\hertz}, the update step could be iterated up to approximately 60 times.

%% file: sections/06_experimental_results.tex
\section{EXPERIMENTAL RESULTS} \label{sec:experimental_results}

We conduct experiments to evaluate the performance of the Equivariant Filter with real-world data.

\subsection{Experimental Setup}
\begin{figure}
    \centering
    \includegraphics[width=\columnwidth]{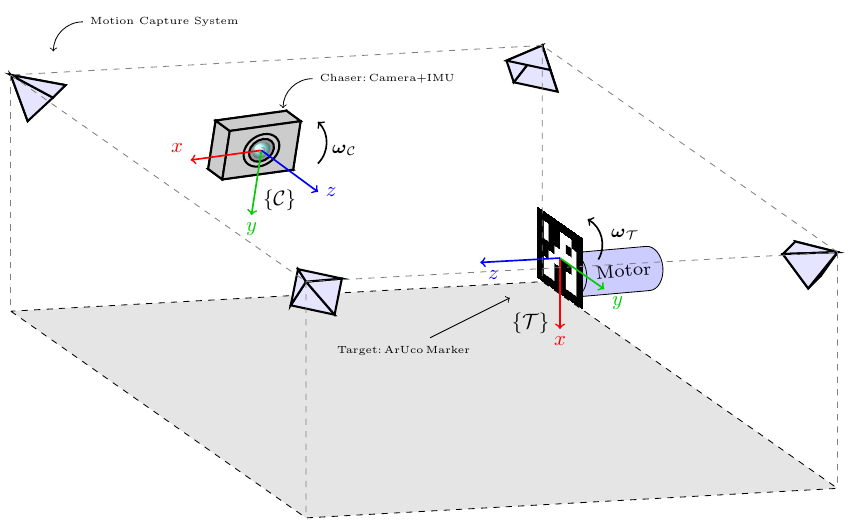}
    \caption{Schematic representation of the experimental setup.}
    \label{fig:experimental_setup}
\end{figure}

To perform the experiments, we set up a fiducial ArUco marker \cite{garridojurado_aruco_2014} on the axis of a DC motor. The marker acts as the target and rotates at a constant angular velocity.
To act as the chaser, we try two different sensor setups. We use an Intel RealSense D435i camera, which acquires images at a frequency of \SI{30}{\hertz} and gyroscope data at \SI{400}{\hertz}. We also use a Prophesee EVK3 event camera, which is capable of high-speed event data output, equivalent to more than \SI{10}{\kilo\hertz} time resolution, while using the RealSense to acquire the gyroscope data. Both are moved freely by hand.
To obtain the ground-truth and evaluate the results, we use an OptiTrack Motion Capture system, running at {\SI{180}{\hertz}}. In Fig.\,{\ref{fig:experimental_setup}}, we provide a schematic of the experimental setup.

\subsection{Acquisition with a Conventional Camera}

We set the target to rotate at a constant angular velocity of approximately {\SI{1}{\radian\per\second}}, or {\SI{57.3}{\degree\per\second}}, about its z-axis, which is aligned with the motor axis. Although the Intel RealSense D435i camera is capable of acquiring at a rate of \SI{30}{\hertz}, in the experiment, the average measurement rate of the ArUco axes, that is, the direction vectors $(\mbf{d}_1, \mbf{d}_2)$, is approximately \SI{15.2}{\hertz}, due to motion blur in some of the images.

\begin{figure}
    \centering
    \begin{minipage}{\columnwidth}
        \centering
        \begin{subfigure}[t]{\textwidth}
            \centering
            \includegraphics[width=\textwidth]{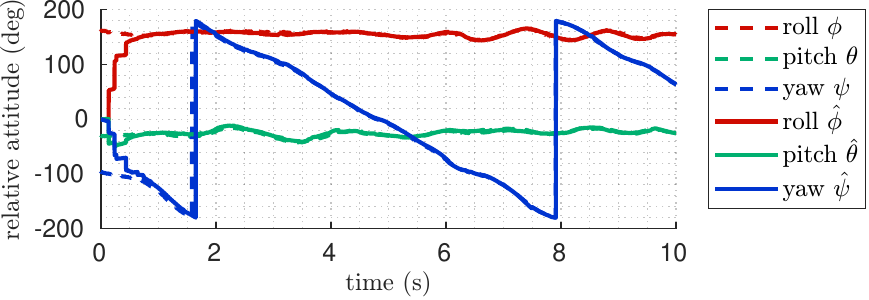}
            \caption{Relative attitude, $\mbf{R}$, expressed in Euler angles.}
            \label{fig:attitude_relative_realsense}
        \end{subfigure}
        \begin{subfigure}[t]{\textwidth}
            \centering
            \includegraphics[width=\textwidth]{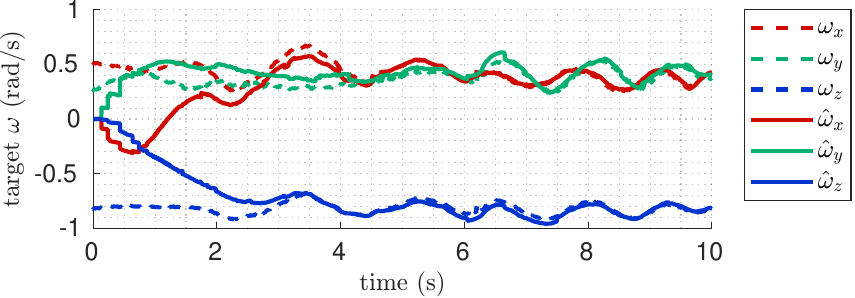} 
            \caption{Target's angular velocity, $\sbf{\omega}$.}
            \label{fig:angular_velocity_target_realsense}
        \end{subfigure}
    \end{minipage}
    \caption{Conventional Camera Experiment: True relative attitude and target's angular velocity $(\mbf{R},\sbf{\omega})$ (dashed) and estimated $(\hat{\mbf{R}},\hat{\sbf{\omega}})$ (solid).}
    \label{fig:manifold_estimates_realsense}
\end{figure}

\begin{figure}
    \centering
    \includegraphics[width=\columnwidth]{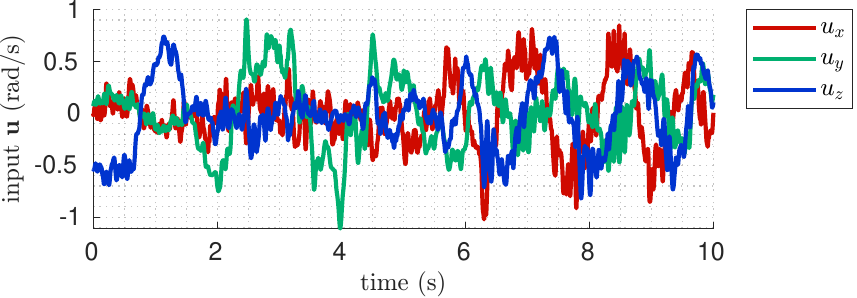}
    \caption{Conventional Camera Experiment: Chaser's angular velocity $\mbf{u}$.}
    \label{fig:angular_velocity_chaser_realsense}
\end{figure}

The estimates of the relative attitude and the angular velocity of the target, expressed in the chaser frame, are displayed in Fig.\;\ref{fig:manifold_estimates_realsense}. The true values are depicted in dashed lines, while the estimates are depicted in solid lines. As in the previous section, the relative attitude is represented in Euler angles. In Fig.\;\ref{fig:angular_velocity_chaser_realsense}, we present the chaser's angular velocity, as measured by the gyroscope in the IMU, which is used as the input $\mbf{u}$ of the EqF.

As the gyroscope data acquisition rate is 26.3 times faster than the direction vectors measurement rate, the results presented here are obtained with the iteration strategy introduced in\;\ref{subsec:effect_of_measurement_rate_in_filter_performance}. Even without this strategy, the filter can still converge with a carefully chosen time step in the discrete-time implementation, though its performance is inferior.

\begin{figure}
    \centering
    \includegraphics[width=\columnwidth]{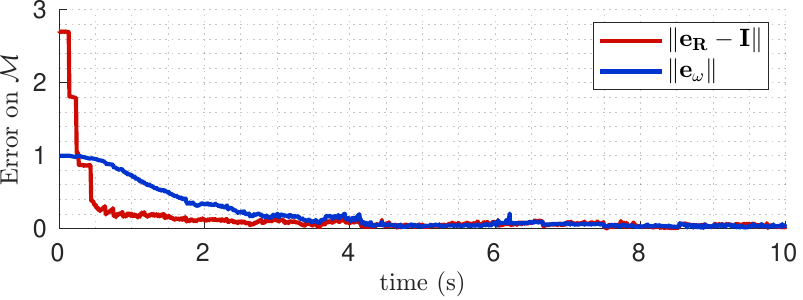}
    \caption{Conventional Camera Experiment: Error norm on the manifold $\mathcal{M}$.}
    \label{fig:error_manifold_realsense}
\end{figure}

In Fig.\;\ref{fig:error_manifold_realsense}, we display the error norm on the manifold. After approximately \SI{4}{\second}, the relative attitude mean error norm is $0.054$, corresponding to a mean error norm of \SI{0.891}{\degree}, \SI{1.331}{\degree}, and \SI{0.934}{\degree} for roll, pitch and yaw, respectively. The target's angular velocity mean error norm is \SI{0.052}{\radian\per\second}, or \SI{2.979}{\degree\per\second}, which is \SI{5.20}{\percent} of the norm of the true target's angular velocity.

\subsection{Acquisition with an Event Camera}

When using a conventional camera, one encounters the problem of motion blur affecting the accuracy of the measurements. As mentioned above, with the target rotating at \SI{1}{\radian\per\second}, the measurement rate is approximately half of its maximum theoretical value, due to blurring in some of the acquired images. If the target is rotating faster or the camera acquisition rate is slower, it will be impossible to acquire data with a conventional camera. To overcome this limitation, we propose the use of an event camera. 

An event camera operates differently from a conventional camera by detecting pixel-level brightness changes, called events. Instead of capturing image frames at a fixed rate, an event camera works in a pixel-by-pixel, independent, and asynchronous manner. By transmitting only changes in brightness, an event camera avoids redundant data transmission, which results in a much higher temporal resolution, in the order of microseconds. Compared to a conventional camera, the Prophesse EVK3 is capable of an equivalent acquisition rate greater than \SI{10}{\kilo\hertz}. 

To demonstrate the potential of an event camera, in this experiment, we set the target to rotate at a constant angular velocity of approximately {\SI{10}{\radian\per\second}} ({\SI{573.0}{\degree\per\second}}). For the results presented here, the direction vectors' measurement rate, acquired by the event camera, is \SI{1}{\kilo\hertz} and the gyroscope acquisition rate is \SI{400}{Hz}. This reverses the situation we had previously, as it is now possible to perform more update steps (with new measurements) than prediction steps. 

\begin{figure}
    \centering
    \begin{minipage}{\columnwidth}
        \centering
        \begin{subfigure}[t]{\textwidth}
            \centering
            \includegraphics[width=\textwidth]{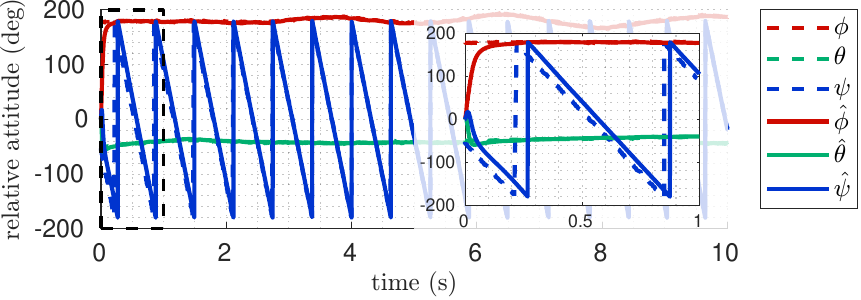}
            \caption{Relative attitude, $\mbf{R}$, expressed in Euler angles.}
            \label{fig:attitude_relative_events}
        \end{subfigure}
        \begin{subfigure}[t]{\textwidth}
            \centering
            \includegraphics[width=\textwidth]{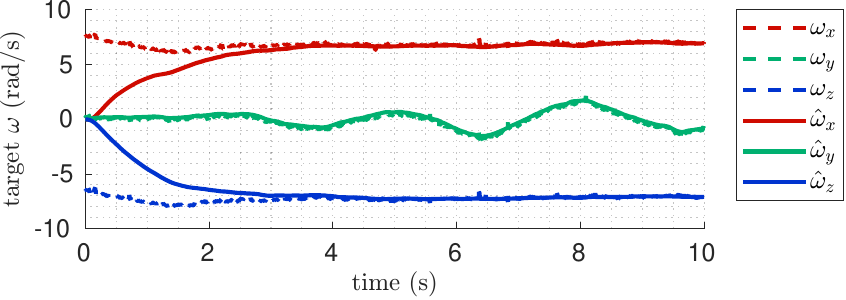} 
            \caption{Target's angular velocity, $\sbf{\omega}$.}
            \label{fig:angular_velocity_target_events}
        \end{subfigure}
    \end{minipage}
    \caption{Event Camera Experiment: True relative attitude and target's angular velocity $(\mbf{R},\sbf{\omega})$ (dashed) and estimated $(\hat{\mbf{R}},\hat{\sbf{\omega}})$ (solid).}
    \label{fig:manifold_estimates_events}
\end{figure}

\begin{figure}
    \centering
    \includegraphics[width=\columnwidth]{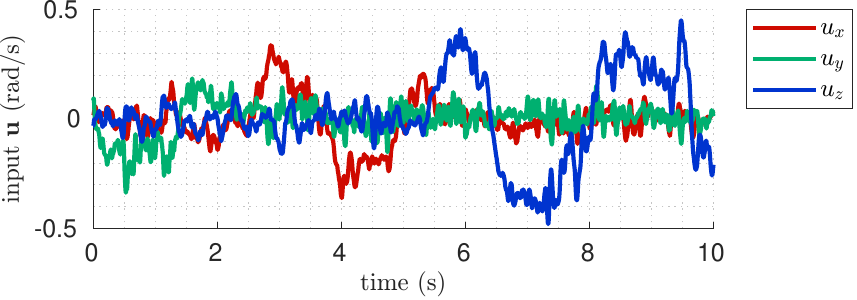}
    \caption{Event Camera Experiment: Chaser's angular velocity $\mbf{u}$.}
    \label{fig:angular_velocity_chaser_events}
\end{figure}

Fig.\;\ref{fig:manifold_estimates_events} shows the estimates of the relative attitude and the angular velocity of the target, expressed in the chaser frame, and Fig.\;\ref{fig:angular_velocity_chaser_events} the chaser's angular velocity.

\begin{figure}
    \centering
    \includegraphics[width=\columnwidth]{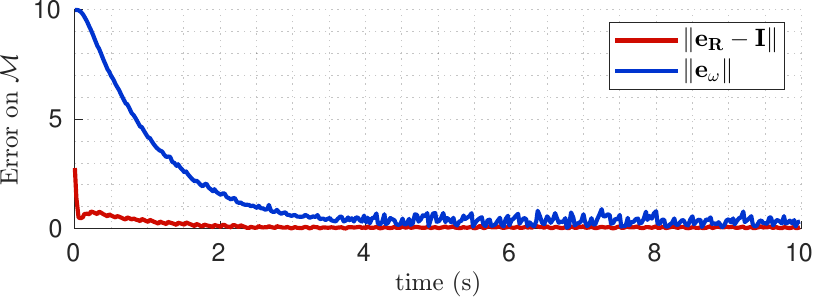}
    \caption{Event Camera Experiment: Error norm on the manifold $\mathcal{M}$.}
    \label{fig:error_manifold_events}
\end{figure}

In Fig.\;\ref{fig:error_manifold_events}, we display the error norm on the manifold. The filter converges after approximately \SI{4}{\second}, with a mean error norm of $0.059$ for the relative attitude, which corresponds to a mean error norm of \SI{1.308}{\degree}, \SI{0.938}{\degree}, and \SI{1.647}{\degree} for roll, pitch and yaw, respectively. The mean of the target's angular velocity error norm of \SI{0.372}{\radian\per\second}, or \SI{21.314}{\degree\per\second}, that is, a relative error of \SI{3.72}{\percent} with respect to the true target's angular velocity norm.

In Tables\;{\ref{tab:convention_event_performance_metrics_relative_attitude}} and {\ref{tab:convention_event_performance_metrics_angular_velocity}}, we present some metrics to analyze the performance of the EqF when using the conventional and event cameras. Table\;{\ref{tab:convention_event_performance_metrics_relative_attitude}} contains the mean error (in\;${\si{\degree}}$) for the relative roll, pitch and yaw angles, in the time interval between 4 and 10\,${\si{\second}}$, the minimum error (in\;${\si{\degree}}$) in the entire experiment, and the time (in\;${\si{\second}}$) that it takes for the filter to achieve a ${\SI{1}{\degree}}$ error in each of the angles. Table\;{\ref{tab:convention_event_performance_metrics_angular_velocity}} contains the absolute and relative mean error norm (in\;${\si{\radian\per\second}}$ and ${\si{\percent}}$, respectively) for the target's angular velocity, in the time interval between 4 and 10\,${\si{\second}}$, the minimum error norm (in\;${\si{\radian\per\second}}$), and the time (in\;${\si{\second}}$) that it takes for the filter to achieve a ${\SI{0.1}{\radian\per\second}}$ error.
{
\sisetup{
  detect-all,
  round-mode=places,
  round-precision=2,
  separate-uncertainty=true,
  output-open-uncertainty = [,
  output-close-uncertainty = ],
}
\begin{table}
    \centering
    \caption{Comparison of conventional and event cameras performance metrics: relative attitude estimation}
    \resizebox{\linewidth}{!}{%
    \begin{tabular}{
        l  
        l  
        S[table-format=1.3]  
        S[scientific-notation=true, table-format=1.2e1]  
        S[table-format=1.2]  
        }
        \toprule
        \multicolumn{2}{c}{} &
        \multicolumn{1}{c}{\shortstack{Mean error (\si{\degree})\\in [4, 10]\,\si{\second}}} &
        \multicolumn{1}{c}{\shortstack{Minimum\\error (\si{\degree})}} &
        \multicolumn{1}{c}{\shortstack{Time to\\1\si{\degree} error (\si{\second})}} \\
        \midrule
        
        \multirow{3}{*}{Conventional} 
          & Roll $\phi$   & 0.89069 & 5.324552e-5 & 0.33193 \\
          & Pitch $\theta$& 1.33103 & 5.637755e-4 & 0.34696 \\
          & Yaw $\psi$    & 0.934339 & 3.921159e-5 & 1.16824 \\
        \midrule
        \multirow{3}{*}{Event}
          & Roll $\phi$   & 1.30769 & 1.8860e-3 & 0.044999 \\
          & Pitch $\theta$& 0.93849 & 3.3304e-3 & 0.026996 \\
          & Yaw $\psi$    & 1.64644 & 3.8088e-4 & 0.420999 \\
        \bottomrule
    \end{tabular}
    }
    \label{tab:convention_event_performance_metrics_relative_attitude}
\end{table}
}
{
\sisetup{
  detect-all,
  round-mode=places,
  round-precision=2,
  separate-uncertainty=true,
  output-open-uncertainty = [,
  output-close-uncertainty = ],
}
\begin{table}
    \centering
    \caption{Comparison of conventional and event cameras performance metrics: target's angular velocity estimation}
    \resizebox{\linewidth}{!}{%
    \begin{tabular}{l
                    S[table-format=1.3]
                    S[table-format=1.3]
                    S[scientific-notation=true, table-format=1.2e1]
                    S[table-format=1.2]}
        \toprule
        {} & 
        {\shortstack{Mean error\\($\si{\radian\per\second}$)\\in [4, 10]\,s}} & 
        {\shortstack{Mean\\error ($\si{\percent}$)\\in [4, 10]\,s}} & 
        {\shortstack{Minimum\\error\\($\si{\radian\per\second}$)}} & 
        {\shortstack{Time ($\si{\second}$) to\\0.1\,$\si{\radian\per\second}$\\error}} \\
        \midrule
        Conventional    & 0.052130 & 5.2130 & 9.75411e-3 & 1.555138 \\
        Event           & 0.372058 & 3.7206 & 3.80754e-2 & 0.984995 \\
        \bottomrule
    \end{tabular}
    }
\label{tab:convention_event_performance_metrics_angular_velocity}
\end{table}

}

\subsection{Discussion and Implications}

These experiments demonstrate the filter's potential in real-world applications for estimating both the relative attitude between a chaser and a target and the target's angular velocity, with different sensors. 

In the first case, with a conventional camera, the update rate is smaller than the prediction rate. As seen in Section\;\ref{subsec:effect_of_measurement_rate_in_filter_performance}, the filter is capable of handling this difference and using the iteration method further improves the performance. 
However, even though the target rotates only at \SI{1}{\radian\per\second}, motion blur deteriorates the acquisition of measurements, resulting in an effective measurement rate of \SI{15.2}{\hertz} and hindering the estimation process. If one were to use a camera with an even lower rate, as those aboard spacecraft, it may be impossible to acquire useful data, rendering this approach impractical.

In the second experiment, where an event camera is used, this issue is overcome, as the camera detects events asynchronously, eliminating the motion blur effect.
With this sensor, we are able to get a high measurement rate, in the order of \SI{1}{\kilo\hertz}, and the filter is able to perform an accurate estimation, even with the target's angular velocity being increased by an order of magnitude to \SI{10}{\radian\per\second}. 


The filter was derived in continuous-time and it assumes that measurements are available at all times. However, in order to run the filter in a computer, one must implement it in discrete-time and the discretization introduces error.
%
Moreover, in a real scenario, the sensor rates can be quite low, which affects the filter's convergence. 
Besides, if the chaser and target have a periodic motion with a frequency that is a multiple of the measurement frequency, then the measurement values would repeat, and the filter would not converge to the true relative attitude and target's angular velocity. 
Nevertheless, this effect can be mitigated if the chaser's motion is not periodic, since each measurement brings new information. 
Furthermore, we have shown that the iterated EqF is able to handle low measurement rates in a discretized programmatic implementation. In addition, this filter demonstrates reduced sensitivity to the gain matrices when compared to an EKF, though tuning is still required in order to achieve optimal performance and the desired transient behavior.

%% file: sections/07_conclusion.tex
\section{CONCLUSION} \label{sec:conclusion}

In this work, we designed and implemented an \revised{Equivariant Filter} for the estimation of the relative attitude between a chaser and a target, and the angular velocity of said target. 
\revised{We showed that the system is equivariant, derived the equivariant lift, and used it to determine the filter dynamics on the group.}
\revised{The error dynamics were derived to design the correction terms and a convergence analysis was provided.}
\revised{To validate performance, we simulated a scenario in which the target exhibits a constant angular velocity and conducted Monte Carlo simulations to ensure a statistically significant result.}
\revised{The EqF was compared against an EKF, showing faster convergence and increased robustness with respect to the choice of state and output gain matrices. Furthermore, we studied the effect of the measurement rate and proposed an iterative strategy to mitigate the negative impact of low-rate measurements.}
\revised{Finally, we demonstrated the approach experimentally. A rotating fiducial ArUco marker was used as the target and both a conventional camera (Intel RealSense D435i) and an event camera (Prophesee EVK3), equipped with an IMU, were employed as the chaser.}
\revised{At the tested target's angular velocity, the conventional camera suffers from motion blur, which degrades the quality and frequency of its measurements. By contrast, the event camera provided accurate measurements, even at a target's angular velocity one order of magnitude higher than in the conventional camera experiment.}
\revised{In both setups, the EqF accurately estimated the relative attitude and the target's angular velocity, thus confirming its effectiveness and potential.}

\revised{In future studies, the approach could be adapted to handle biased measurements of the chaser's angular velocity, by simultaneously estimating the gyroscope bias, which would increase the filter's robustness. The method could also be extended to estimate the relative pose between the spacecraft, thereby broadening its applicability to complex proximity and rendezvous missions.}

%% file: sections/99_appendix.tex
\section*{APPENDIX}

\label{app:proof_equivariant_lift}
We want to prove that the lift \eqref{eq:equivariant_lift} is, in fact, equivariant.

\begin{proof}
If the lift is equivariant, then the conditions \eqref{eq:equivariant_lift_condition_1} and \eqref{eq:equivariant_lift_condition_2} must be verified. First, we calculate
\begin{equation}
\begin{split}
    D&_{{(\mbf{Q}, \mbf{q})|_{(\mbf{I}, \mbf{0})}}}\phi_{(\mbf{R}, \sbf{\omega})}(\mbf{Q},\mbf{q})[(\Lambda_\mbf{Q},\Lambda_\mbf{q})]\\
    &=\lim_{t\to0} \frac{1}{t}\Bigl( \phi_{(\mbf{R},\sbf{\omega})}(\mbf{Q} + t\Lambda_{\mbf{Q}},\mbf{q} + t\Lambda_{\mbf{q}}) \\
    & \hspace{30mm}- \phi_{(\mbf{R},\sbf{\omega})}(\mbf{Q}, \mbf{q}) \Bigr)_{{|(\mbf{Q}, \mbf{q})=(\mbf{I}, \mbf{0})}}\\
    %
    %
    &=\bigl( \mbf{R}(\mbf{u} - \sbf{\omega} + \mbf{v})^{\wedge},\, \mbf{a} + (\sbf{\omega} + \mbf{w})\times \mbf{u} \bigr)\\
    &=\bar{f}((\mbf{R}, \sbf{\omega}), (\mbf{u},\mbf{a},\mbf{v},\mbf{w}))\,,
\end{split}
\end{equation}
which means that the condition \eqref{eq:equivariant_lift_condition_1} is verified. To prove the condition \eqref{eq:equivariant_lift_condition_2}, let us calculate the left-hand side of the expression, which is given by 
\begin{equation}
\begin{split}
    \text{Ad}&_{(\mbf{Q}, \mbf{q})^{-1}} (\Lambda_\mbf{Q},\Lambda_\mbf{q}) = \text{Ad}_{(\mbf{Q}^\T, -\mbf{Q}^\T\mbf{q})} (\Lambda_\mbf{Q},\Lambda_\mbf{q})\\
    &= \bigl( \mbf{Q}^\T \Lambda_\mbf{Q} \mbf{Q}, \mbf{Q}^\T \Lambda_\mbf{Q} \mbf{q} + \mbf{Q}^\T \Lambda_\mbf{q}  \bigr) \\
    &=  \bigl( \mbf{Q}^\T (\mbf{u} - \sbf{\omega} + \mbf{v})^{\wedge} \mbf{Q},\,  \mbf{Q}^\T (\mbf{u} - \sbf{\omega} + \mbf{v})^{\wedge} \mbf{q}  \\
    & \hspace{20mm}+ \mbf{Q}^\T (-\mbf{a} + \mbf{u}\times\mbf{w} + \sbf{\omega}\times \mbf{v})\bigr)\,.\\
\end{split}
\end{equation}
Now, we calculate the right-hand side of \eqref{eq:equivariant_lift_condition_2}, 
\begin{equation}
    \begin{split}
        \Lambda&\bigl(\phi_{(\mbf{Q}, \mbf{q})}(\mbf{R}, \sbf{\omega}), \psi_{(\mbf{Q}, \mbf{q})}(\mbf{u}, \mbf{a}, \mbf{v}, \mbf{w})\bigr)\\
        =& \bigl( (\mbf{Q}^\T \mbf{u} - \mbf{Q}^\T(\sbf{\omega} - \mbf{q}) + \mbf{Q}^\T (\mbf{v}-\mbf{q}))^{\wedge},\,-\mbf{Q}^\T \mbf{a} \\
        &\hspace{22mm} + (\mbf{Q}^\T \mbf{u}) \times (\mbf{Q}^\T (\mbf{w} + \mbf{q})) \\
        &\hspace{22mm} + (\mbf{Q}^\T (\sbf{\omega} - \mbf{q})) \times (\mbf{Q}^\T (\mbf{v} -\mbf{q}))\bigr)\\
        =&\bigl( \mbf{Q}^\T (\mbf{u} - \sbf{\omega} + \mbf{v})^{\wedge} \mbf{Q},\,\mbf{Q}^\T (\mbf{u} - \sbf{\omega} + \mbf{v})^{\wedge} \mbf{q} \\
        &\hspace{22mm}+ \mbf{Q}^\T (-\mbf{a} + \mbf{u}\times\mbf{w} + \sbf{\omega}\times \mbf{v})\bigr)\,.\\
    \end{split}
\end{equation}
We conclude that condition the \eqref{eq:equivariant_lift_condition_2} is also verified. Thus, the proposed lift is equivariant.
\end{proof}